\documentclass[11pt]{article}
\usepackage{graphicx}
\usepackage{amsmath}
\usepackage{amssymb}
\usepackage{amsbsy}
\newcommand{\bit}{\begin{itemize}}
\newcommand{\eit}{\end{itemize}}

\def\benu{\begin{enumerate}}
\def\eenu{\end{enumerate}}
\def\noi{\noindent}
\def\btab{\begin{tabbing}}
\def\etab{\end{tabbing}}

\def\bit{\begin{itemize}}
\def\eit{\end{itemize}}
\def\beq{\begin{equation}}
\def\eeq{\end{equation}}
\def\bec{\begin{center}}
\def\eec{\end{center}}
\def\btable{\begin{tabular}}
\def\etable{\end{tabular}}
\def\beqr{\begin{eqnarray}}
\def\eeqr{\end{eqnarray}}
\def\rarw{\rightarrow}

\def\om{\omega}
\def\gm{\gamma}
\def\Gm{\Gamma}

\def\lm{\lambda}
\def\eps{\epsilon}

\def\alp{\alpha}
\def\bt{\beta}

\def\Dl{\Delta}
\def\sg{\sigma}
\def\Om{\Omega}
\def\rarw{\rightarrow}
\def\del{\partial}

\def\half{\frac{1}{2}}

\def\btab{\begin{tabbing}}
\def\etab{\end{tabbing}}
\def\beqrs{\begin{eqnarray*}}
\def\eeqrs{\end{eqnarray*}}
\def\noi{\noindent}
\def\lan{\langle}
\def\ran{\rangle}

 \fontfamily{ptm}\selectfont
\usepackage{mathptmx}           % this gives slanted Times symbols for math

\date{}

\begin{document}

%\tableofcontents
%\clearpage

\title{Spectral Brilliance of Channeling Radiation at the ASTA Photoinjector}

\author{Tanaji Sen\footnote{tsen@fnal.gov} \\ Fermi National Laboratory \\ 
Batavia, IL 60510 \\ 
\mbox{} \\
Christopher Lynn \\ Swarthmore College, PA 19081}

\maketitle

\begin{abstract}
We study channeling radiation from electron beams with energies
under 100 MeV. We introduce a phenomenological model of dechanneling, correct
non-radiative transition rates from thermal scattering, and discuss in detail
the population dynamics in low order bound states. 
These are used to revisit the X-ray properties measured at the ELBE facility
in Forschungszentrum Dresden-Rosenstock (FZDR), extract parameters for 
dechanneling states, and obtain satisfactory
agreement with measured photon yields. The importance of rechanneling 
phenomena in thick crystals is emphasized. 
The model is then used to calculate the expected X-ray energies, linewidths
and brilliance for forthcoming channeling radiation experiments at Fermilab's
 ASTA photoinjector. 
\end{abstract}
%\keywords{Channeling Radiation; Photoinjector; Spectral Brilliance.}

%\ccode{PACS numbers:}

\section{Introduction}

Channeling radiation offers the promise of a quasi-monochromatic and tunable 
X-ray source with electron beams
of moderate energies (tens of MeV) passing through a thin crystal.
This radiation has been experimentally observed at several laboratories and many
of the experimental features are well understood from theoretical considerations.
Reviews can be found in several publications, see e.g. 
Refs.~\cite{Andersen_rev_83, Saenz, Uggerhoj}. 

Channeling and channeling radiation experiments have a long history at Fermilab,
see e.g. Ref.~\cite{Carrigan_03}. Those were carried out at the A0 photoinjector which
had a maximum beam energy of about 15 MeV. A new photoinjector ASTA is being 
commissioned at Fermilab, which will use an 
L-band (1.3 GHz) linac to generate beams with energies initially in the range 
20-50 MeV and later to 300 MeV and higher with the addition of one or 
more ILC style cryomodules \cite{Piot_13}. Channeling radiation experiments 
with beams in the lower energy range have been planned and descriptions of the 
planned
experiments can be found in Refs.~\cite{Brau, Piot_12}. The goal is to generate X-ray 
beams with high average brilliance using low emittance electron beams with the aim
of increasing the brilliance by about six orders of magnitude over that obtained 
with channeling experiments conducted at FZDR's  ELBE linac \cite{Wagner}.  
Once demonstrated, 
compact X-ray sources from channeling radiation can be designed and built with X-band linacs. 

In this report we revisit the theoretical model for channeling radiation with the 
aim of improving the calculation of the
X-ray intensity. We compare the calculations of the revised model with the measurements of 
previous experiments at ELBE
and find better agreement of the photon yields with the experimental values shown
in ref.~\cite{Azadegan_thesis}. We then use this model to calculate the
expected photon yields and X-ray brilliance at ASTA.

\section{Theoretical model}

In the case of planar channeling, the particle motion in its rest frame can
be well approximated by motion in a single transverse direction (here
$x$) orthogonal to the plane. For particle energies below 100 MeV, the X-ray 
energy spectrum is discrete and  the radiation
 is best understood as emitted during transitions between the discrete bound states in the
crystal potential and requires a quantum mechanical treatment. The Schroedinger equation for the electron wave function $\psi(x)$ in the particle rest frame is
\beq
[-\frac{\hbar^2}{2m_e \gm} \frac{\del^2}{\del x^2} + V(x)]\psi(x) = E_{\perp}\psi(x)
\eeq
Here $\gm = 1/\sqrt{1-(v/c)^2}$ is the usual kinematic factor related to the 
velocity $v$ and
$V(x)$ is the one dimensional continuum potential obtained by averaging 
the
three dimensional atomic potential along the orthogonal directions $(y,z)$.
Taking into account the lattice periodicity, the potential can be expanded as a
Fourier series
\beq
V(x) = \sum_{n=-\infty}^{\infty} V_n \exp[i n g x]
\label{eq: Vx}
\eeq
Here $g=2\pi/d_p$ is the lattice spacing in reciprocal space while $d_p$ is the
lattice spacing in direct space. 
The Fourier coefficients $V_n$ are typically obtained from expanding the
electron form factor $f_{el}(4\pi s)$ (defined in Eq.(\ref{eq: f_el})) into a sum of four
Gaussians with four Doyle-Turner coefficients $(a^{DT}, b^{DT})$ 
\cite{Doyle-Turner}. We use here instead six coefficients as used in 
Ref.~\cite{Chouffani} which extends the range of validity of the approximation
from $s\le 2$\AA$^{-1}$ to $s\le 6$\AA$^{-1}$ for planar channeling. 
\beq
V_n = -\frac{2\pi}{V_c}a_0^2 (\frac{e^2}{a_0})e^{-M(\vec{g})}
\sum_j e^{i \vec{g}\cdot \vec{r}_j} \sum_{i=1}^6 a_i^{DT}
\exp[-\frac{b_i^{DT}}{16\pi^2}(ng)^2]
\label{eq: V_n}
\eeq
Here $\vec{g}$ is the reciprocal lattice vector, $V_c$ is the volume of the 
unit cell, $a_0$ is the Bohr radius, $\vec{r}_j$ are the coordinates of the $j$th
atom in the unit cell and $M(\vec{g}) = \half g^2 \lan u_{th}^2  \ran$ is the 
Debye-Waller factor describing thermal vibrations with mean squared amplitude
$\lan u_{th}^2\ran$, assumed to be the same for all atoms. 

The wave function solution for the periodic potential is given in terms of 
Bloch waves
\beq
\psi(x) = \frac{e^{i k x}}{\sqrt{d_p}} \sum_{n=-\infty}^{\infty} c_n \exp[i n g x]
\label{eq: psi_Bloch}
\eeq
where $k$ is the electron wave number. 
In practice, the Fourier expansion for the potential and the wave function
is limited to a finite number of modes $M$, in the cases considered here
$M=20$. Substitution of Eqs. (\ref{eq: Vx}) and (\ref{eq: psi_Bloch})
into the Schroedinger equation reduces it to an eigenvalue problem with
a matrix $A$ whose components are \cite{Azadegan_PRB}
\beqr
A_{nn} & = & \frac{\hbar^2}{2m_e \gm}(k + ng)^2 + V_0 \nonumber \\
A_{nm} & = & V_{n-m}, \;\;\; n \ne m 
\eeqr
Solutions to the eigenvalue problem results in the eigen-energies $\eps_n$ and the
coefficients $c_n$ determining the wavefunctions $\psi(x)$.

\subsection{Radiative transitions}

Radiative transitions from one state to another lead to photon emission and the 
transition rates are given by 
Fermi's golden rule which states that the transition rate per unit solid angle, 
per length of traversal into the crystal and per unit photon energy
is proportional to the
matrix element of the transition operator between the states, i.e.
\beq
\frac{d^3 N}{d\Om dz dE_{\gm}}(n \rarw m) \propto |\lan \psi_m|\frac{d}{dx}|\psi_n\ran|^2 P_n(z)
\eeq
where $d/dx$ corresponds to the dipole operator and $P_n(z)$ is the 
probability of occupation in the state $|\psi_n \ran$ at a distance $z$
into the crystal.

Applying this rule yields the differential energy angular spectrum from a state $n$ to state $m$ 
as \cite{Andersen_83,Kephart, Genz_96, Azadegan_PRB}
\beqr
\frac{d^2N}{d\Om dE_{\gm}} & = & \frac{\alp_f \lm_C^2}{\pi^{5/2}\hbar c}
2 \gm^2 (\eps_n - \eps_m)|\langle \psi_m| \frac{d}{dx}|\psi_n\rangle|^2
\nonumber \\
& & \mbox{} \!\!\!\!\!\!\!\!\!  \times \!\!\!\!\!\!
\int_0^d dz e^{[-\mu(E_{\gm})(d-z)]}P_n(z) \!\!\! \int_0^{\infty} dt
\!\!\! \frac{t^{-1/2}(1+2\alp^2 t)
(\Gm_T/2) e^{-t}}{[(1+2\alp^2 t)E_{\gm}-E_0]^2+[(1+2\alp^2 t)(\Gm_T/2)]^2}
\label{eq: d2N_domdE}
\eeqr
where $\alp_f$ is the fine structure constant, $\lm_C$ is the Compton wavelength of the electron,
 $\eps_n,\eps_m$ are the energies of the states $n, m$ respectively, 
$d$ is the crystal thickness, 
$\mu(E_{\gm})$ is the energy dependent photon absorption coefficient, $E_{\gm}$ is the X-ray energy at the angle of observation,
$E_0$ is the X-ray energy at zero angle, $\alp = \gm \theta_{MS,ch}$ where 
$\theta_{MS,ch}$ is the multiple scattering angle during channeling and
$\Gm_T$ is the total linewidth of the transition $n\rarw m$ line. 
From this the differential angular spectrum is found from
\beq
\frac{dN}{d\Om} = \int_{E_{\gm}-\Gm_T/2}^{E_{\gm}+\Gm_T/2} \frac{d^2N}{d\Om dE_{\gm}}
dE_{\gm}
\eeq
where the integration is done over the linewidth of the spectral line with its
peak at $E_{\gm}$.
These transitions only occur between states of opposite parity because the 
dipole
transition matrix element is non-zero only between these states. 
The dipole operator is of odd-parity and the bound states $|\psi_n\rangle$ in 
planar channeling are states of definite parity: even parity for n even and odd
for n odd. 

With the wave function defined in terms of the Fourier coefficients $c_n$,
the dipole matrix elements between two states is given by
\beq
\langle \psi_m| \frac{d}{dx}|\psi_n\rangle = i \frac{2\pi}{d_p}
\sum_{j=-M}^M (jg+k) (c_j^m)^* c_j^n
\eeq
where $d_p$ is the inter-planar distance and $c_j^m, c_j^n$ are the coefficients
in the Bloch wave expansion (see Eq.(\ref{eq: psi_Bloch})) of $\psi_m,\psi_n$
respectively.

If the eigen-energies of the two states involved in the transition are $\eps_m, \eps_n$, the
energy of a photon emitted at zero angle and emitted at angle $\theta$ with the beam direction are
given by
\beqr
E_{\gm}(0) & = & 2\gm^2(\eps_n - \eps_m) \nonumber \\
E_{\gm}(\theta) & = & \frac{E_{\gm}(0)}{1 - \beta \cos\theta} \approx
\frac{E_{\gm}(0)}{1 + \gm^2 \theta^2}, \;\;\; \theta \ll 1
\label{eq: E_gamma}
\eeqr

\subsection{Non-Radiative transitions} \label{subsec: NonRad}

In addition to the radiative transitions, electrons can also change energy by non-radiative 
transitions which we discuss here. 
Electrons in the lower bound states are
closer to the atomic nuclei and can change energy due to thermal scattering
with the vibrational motion of the atoms. This energy exchange can
lead to a change of the wave vectors within the same Brillouin zone 
(intra-band scattering) or even transfer them to different energy states (inter-band
scattering). This electron-phonon scattering is the dominant contribution
to the non-radiative transitions that change the populations of the states;
a relatively smaller  contribution is played by the
 electron  scattering off atomic electrons. The transition probability due to thermal
scattering that an electron will move from state $k, \psi_n$ with momentum
$k$ to state $k^{'}, \psi_m$ with a different momentum $k^{'}$ is given
by a transition rate per unit length $W_{kn, k^{'}m}$ where
\beq
W_{kn, k^{'}m} = W_{k^{'}m, k n} = \frac{2}{\hbar v}| \lan k^{'}, \psi_m| V^I |
k, \psi_n \ran
\label{eq: Wmn}
\eeq
where the potential $V^I$ describes the inelastic scattering. It is the
imaginary part of a complex potential with the real part being the continuum
potential $V(x)$ which describes the elastic scattering. 
Intra-band scattering is described by $m=n$ while inter-band scattering 
has $m \ne n$. 
Calculation shows that the variation of the rate
within an energy band (Brillouin zone) is smaller than the variation between
bands. For clarity of notation, we will drop the momentum
indices from $W$ in the following but they are understood to be present. 

Similar to the real potential, the imaginary potential can also be expanded in 
a Fourier series as
\beq
V^I({\bf r})  =  \sum_{\bf g} V_{\bf g}^I \exp[i {\bf g}\cdot{\bf r}]
=  \sum_n V_{n g}^I \exp[i n g \; \hat{\bf g}\cdot {\bf r}]
\eeq
where $\hat{\bf g}$ is the unit reciprocal lattice vector.

We briefly summarize the procedure for calculating the imaginary potential and hence 
the Fourier coefficients $ V_{n g}^I$, following 
the method in Ref.~\cite{Chouffani}. 
As is done in solving for the energy eigenvalues, the incident and scattered wave
 functions of the electron
are represented as sums of Bloch functions with the sums extending
over many reciprocal lattice planes. In general thermal scattering occurs in
all three directions, hence a three dimensional formalism is necessary.
The incident and scattered wave functions can be written as
\beqr
\psi_{inc}({\bf r}) & = & \sum_{n} a_n \exp[i({\bf k}_0 + {\bf g}_n)
\cdot {\bf r} \nonumber \\
\psi_{scat}({\bf r}) & = & \sum_{n} \sum_j a_n \gm 
f_{el}({\bf q}- {\bf g}_n) \exp[i({\bf q}- {\bf g}_n)\cdot {\bf r}_j]
\exp[i {\bf k}\cdot {\bf r}] 
\label{eq: elec_wave}
\eeqr
Here $a_n$ are the coefficients in the expansion, the sum over $n$ extends over 
reciprocal lattice planes,
${\bf g}_n$ are the reciprocal lattice vectors while the sum over $j$
extends over the atoms in the crystal. ${\bf k}_0, {\bf k}$ are the incident and outgoing wave vectors
of the electron and  ${\bf q} = {\bf k} - {\bf k}_0$. $f_{el}$ is the electron scattering form factor given by
\beq
f_{el}({\bf q}) = \frac{m}{2\pi \hbar^2} \int V({\bf r})
\exp[-i {\bf q}\cdot {\bf r}] d{\bf r}
\label{eq: f_el}
\eeq
Here $V({\bf r})$ is the real part of the atomic potential. 

The transition rate is found from the intensity of the thermally diffuse 
scattering 
\beq
W =\frac{v}{V_c}\int d\Om \;  I_{diff}
\label{eq: W_int}
\eeq
where $d\Om$ is the solid angle into which the particle is scattered, 
$V_c$ is the volume of the crystal. The diffuse scattering intensity
 is related to the total scattering intensity via
\beq
I_{total} \equiv \lan \psi_{scat}|\psi_{scat}\ran  = I_{diff} + I_{Bragg} 
\eeq
Here $I_{Bragg}$ is the Bragg scattering or the coherent scattering intensity.
Bragg scattering does not generally lead to energy exchange between electrons
and the atoms but instead modulates the amplitude of the electron
intensity in the crystal \cite{Hall_Hirsch}. It occurs even when the atoms
are stationary, so the diffuse inelastic scattering intensity is found by
extracting only the contribution from the thermally induced vibrations.

In evaluating the scattered intensity, the instantaneous position can be 
represented as
${\bf r}_j(t) =  {\bf R}_j + {\bf u}_j(t)$
where ${\bf R}_j$ is the stationary equilibrium position of the
$j$th atom while ${\bf u}_j(t)$ represents the time dependent
position due to thermal vibrations. Then a thermal averaging is performed assuming that the thermal vibrations are isotropic.
The terms in the total scattered intensity $I_{total}$ that are independent of the atomic coordinates ${\bf R}$
define the incoherent thermally diffuse scattering
\beqr
\lan I_{diff} \ran & = & N_{at}\gm^2 \sum_m \sum_n a_m^* a_n 
f_{el}({\bf q}- {\bf g}_m)f_{el}({\bf q}- {\bf g}_n) \nonumber \\
& & \times \left\{
\exp[-\half u_{th}^2(({\bf g}_m-{\bf g}_n)^2)] - 
\exp[-\half u_{th}^2(({\bf q}-{\bf g}_m)^2+({\bf q}-{\bf g}_n)^2)]
\right\} 
\eeqr
Here $\lan \mbox{} \ran$ represents the thermal average, 
$\lan {\bf u}^2\ran = u_{th}^2$, and $N_{at}$ is the number of atoms in the 
crystal. This incoherent intensity would vanish in the absence of thermal 
vibrations. 

Equating the two expressions for the transition rates in Eq.(\ref{eq: Wmn}) and Eq.(\ref{eq: W_int})
leads to an expression for the Fourier coefficients
\beqr
V_{\bf g}^I  & = & \frac{\hbar \bt c}{2 V_c} 
\frac{\gm^2 N_{at}}{k_0^2} \int \int q dq d\phi f_{el}({\bf q})
f_{el}({\bf q}-{\bf g}) \nonumber \\
& & \times \left[ \exp(-\half u_{th}^2g^2) - 
 \exp(-\half u_{th}^2(q^2 + ({\bf q}-{\bf g})^2)) \right]
\eeqr
The element of solid angle in this integral is written as
$d\Om=\sin\theta d\theta d\phi = (q dq/k_0^2) d\phi$. The second equality follows 
from ${\bf q} = {\bf k} - {\bf k}_0$, hence $q dq = k_0^2\sin\theta d\theta$ when
$|{\bf k}| \approx |{\bf k}_0|$.
Using the  Doyle-Turner like expansions.for the real potential, the electronic form factors can be written as
\beq
f_{el}({\bf q})  =  \sum_i a_i^{DT} 
\exp[-(b_i^{DT} + 8\pi^2 u_{th}^2)q^2/(16\pi^2)] = 
 \sum_i a_i^{DT} \exp[- B_i q^2]
\eeq
where 
\[ B_i = \frac{1}{16\pi^2}b_i^{DT}  + \half u_{th}^2 
\]

Performing the integrations, the expression for the Fourier coefficients of
the imaginary potential is
\beqr
V_{\bf g}^I & = & \frac{\hbar^3}{2\bt m_e^2 c} \frac{N_{at}}{V_c} \exp[-(\half u_{th}^2 g^2)]
\sum_i\sum_j a_i^{DT} a_j^{DT}
\nonumber \\
& & \left[ \frac{1}{B_i+B_j}\exp[-\frac{B_i B_j}{B_i+B_j}g^2]
- \frac{1}{B_i+B_j+ u_{th}^2}
\exp[-(\frac{B_i B_j - u_{th}^4/4}
{B_i + B_j + u_{th}^2})g^2] \right] \nonumber \\
 & &   \mbox{}
\label{eq: Vimag}
\eeqr
Again, we note that the coefficients $V_{\bf g}^I$ vanish in the absence
of thermal vibrations $u_{th}=0$. This potential depends on the particle
energy only through $\bt$, hence this potential is nearly independent of
particle energy for relativistic electrons. 
This expression differs from the incorrect expression (Eq.(A20)) in 
Ref.~\cite{Chouffani}.
The resulting imaginary potential turns out to have a smaller magnitude
and the opposite sign to the imaginary potential used in the numerical modeling
for the ELBE experiments, e.g Eq.(11) in reference \cite{Azadegan_PRB}.

From the numerical 
calculations, the transition rates $W_{m, n}$ are found to obey the 
approximate selection rule that only same parity transitions are allowed, 
i.e. $|m- n|=even$. The odd parity transitions are non-zero but small.

The probability $P_{n}$ of a state  $\psi_n$ changes as the electron propagates 
through
the crystal. The rate of change is determined by the transition rates 
$W_{m,n}$  as 
\beq
\frac{dP_{n}}{dz} = \sum_m W_{m, n}[P_{m}(z)-P_{n}(z)]
\label{eq: Pn_orig}
\eeq
where $W_{m, n}$ is the transition probability from a state $|\psi_m\rangle$
to state $|\psi_n\rangle$. The first term in the sum corresponds to entering the
state $|\psi_n\rangle$ from other states while the second term corresponds to 
electrons leaving that state. In this model, equilibrium populations, i.e. $dP/dz=0$ 
are reached when the populations in all states are equal $P_m=P_n$. 

\subsection{Dechanneling}

If the electron is scattered into a free state, it is possible
that the electron will remain in a free state and not be scattered back into
a bound state while propagating through the crystal. In addition, multiple 
scattering can move electrons from lower  states to higher 
states and effectively remove electrons from contributing to the radiation
yield. This enhanced dechanneling can be taken into account phenomenologically in 
the above model by removing 
those electrons scattered into free state above a certain energy
from contributing to the photon yield. Thus the above equation would be
modified to a set of two equations. If $n_f$ denotes the free state at which
electrons are dechanneled and do not scatter back into the bound states, then
the propagation of the probabilities are given by
\beqr
\frac{dP_{n}}{dz} & = & \sum_{m < n_f}W_{m,n} P_{m} - \sum_{m=1}^M W_{m,n} 
P_{n} ; \;\;\;\; n < n_f \label{eq: Pn_mod_1}\\
\frac{dP_{n}}{dz} & = & \sum_{m=1}^M W_{m,n} P_{m} - \sum_{m \ge n_f} 
W_{m,n} P_{n}
; \;\;\;\; n \ge n_f \label{eq: Pn_mod_2}
\eeqr
The first term in Eq (\ref{eq: Pn_mod_1}) restricts the electrons entering 
state $n$ to only those from states below $n_f$ while the second term allows 
the escape of 
electrons from this state to all states. Similarly for states at and above $n_f$, 
electrons can enter from all states (1st term in Eq.(\ref{eq: Pn_mod_2})) but can 
only escape to states above $n_f$.
These set of equations conserve population, i.e. 
$ (d/dz)\sum_{n=1}^M P_n(z) = 0$, as they should. In this model there are 
no equilibrium solutions except for those states which are depopulated at 
the beginning of the crystal and remain so. For the other states, the 
asymptotic 
solutions in this model at large $z$ are given by $ dP_n/dz < 0$ for $n< n_f$ 
and $dP_n/dz \ge 0$ for $n \ge n_f$. 

A complete theory would have transition rates from multiple scattering in the 
model, but in its absence we will use
experimental data to find the best value for $n_f$. In the cases we will consider,
$n_f$ is found to be determined primarily by the crystal thickness and not 
by the electron energy. We expect that $n_f$ will increase with crystal thickness
to reflect the higher rechanneling probability that electrons will scatter into 
bound states
with increasing thickness. This was observed to be the case in 
experiments at the Darmstadt linac \cite{Genz_96} and our fits to the
experimental data from the ELBE linac will also show this to be the case.

A proper treatment of dechanneling at low energies using a quantum mechanical
treatment still seems to be lacking. However, 
when the electron energy is high enough that channeling radiation can be treated
classically, then a Fokker-Planck treatment of the diffusive motion has been used
to describe dechanneling. Results from one such analysis and comparison with
experiments using 855 MeV electrons were reported in Ref.~\cite{Backe_08}. 
A brief
survey of dechanneling phenomena at high energies for negatively and positively 
charged particles was reported in Ref.~\cite{Carrigan_08}.

\subsection{Line width and length scales}

The finite lifetime of quantum states, Bloch wave broadening of each energy band due to the
variation of transverse momenta, and multiple
scattering are the dominant sources of line broadening in the regime of our interest. Other less
significant sources are the Doppler broadening due to emission at non-zero angles, electron beam energy
spread and detector resolution. Here we
will consider the  dominant effects and the length scales associated with line 
broadening effects, more complete discussions can be found in
Refs.~\cite{Chouffani, Azadegan_PRB}.

{\it Coherence length}: This is a measure of the length over which a radiating electron stays in
phase and it determines the lifetime of the bound states.
Thermal (phonon) scattering, atomic electron (plasmon) scattering and other incoherent scattering effects
change the phase of the initial wave function of the electrons and they lose 
coherence. These scattering
effects can be described by the imaginary potential discussed above. The 
coherence length $L_{coh}$ for transitions between two states $n, m$  is given 
by 
\beq
\frac{1}{L_{coh}} = \frac{1}{l_n} +  \frac{1}{l_m} , \;\;\;\;
l_m = \frac{\hbar \bt c}{2 \langle V_m^I \rangle}
\eeq
where $ \langle V_m^I \rangle $ is the expectation value of the imaginary 
part of the complex potential in the state $m$. 
The line width due to this finite lifetime of each state is given by
\beq
\Gm_{coh} = \frac{2\gm^2 \hbar \bt c}{L_{coh}}
\eeq
The correction to the imaginary potential
discussed in the previous section makes it smaller and hence the calculated 
coherence lengths are larger than those reported in Ref.~\cite{Azadegan_PRB}.

{\it Bloch wave broadening}: Each energy band has a finite width due to the spread of
wave vectors within each Brillouin zone. Hence the energy spread from transitions 
between states $n$ and $m$ is given by
\beq
\Gm_{BW} = 2\gm^2(|\eps_n^{k_{\perp}=0} - \eps_n^{k_{\perp}=g/2}| + 
|\eps_m^{k_{\perp}=0} - \eps_m^{k_{\perp}=g/2}|)
\eeq
where $g$ is the reciprocal lattice spacing and $k_{\perp}$ is the transverse 
component (here $k_x$) of the wave vector. The width is larger for the higher 
bound states because of their larger range of transverse momentum. The
relative importance of Bloch wave broadening increases with energy as the
number of bound states increases. 

{\it Multiple scattering  of channeled particles}: This contributes to the line
width by changing the angle of the scattered electron and hence the
angle at which photons are emitted and their energy. 
For particle scattering in an amorphous medium, the rms scattering angle is
given by \cite{PDG}
\beq
\theta_{MS} = \frac{13.6}{E_e {\rm [MeV]}}\sqrt{\frac{d}{L_r}}[1 + 0.038
\log\frac{d}{L_r}]
\eeq
where $d$ is the thickness and  $L_r$ is the radiation length, the 
length over which an electron loses $1/e$ of its initial energy. This
expression is considered to be accurate to about 10\% for 
thickness down to 0.001 $L_r$ \cite{PDG}. 
Experimental estimates show that the
scattering angle of channeled particles in a crystal is less than that in
an amorphous medium. Estimates of the scaling between the rms multiple scattering
angles during channeling and in an amorphous crystal are in the range 
$ \theta_{MS,ch} \simeq (0.22 - 0.56) \theta_{MS}$
\cite{Chouffani, Azadegan_PRB}. 
The range of values in the numerical coefficient depends on the crystal 
thickness, smaller values for
larger thickness, but is nearly independent of the beam energy. 

The change in the angle of emission Doppler shifts the photon energy, with its
energy given by the second equation in Eq.(\ref{eq: E_gamma}).
The mean photon energy is found by
averaging the angle dependent energy over the distribution of multiple 
scattering angles, assumed to be Gaussian. Thus
\beq
\lan E_{\gm}\ran \equiv \frac{1}{\sqrt{2\pi}\theta_{MS,ch}}
\int E_{\gm}(\theta) \exp[-\frac{\theta^2}{2\theta_{MS,ch}^2}] d\theta
\label{eq: Eav_MS}
\eeq
while the rms width of this distribution may be taken as a measure of the
linewidth due to multiple scattering,
\beq
\Gm_{MS} = \sqrt{\lan E_{\gm}^2\ran - \lan E_{\gm}\ran^2}
\eeq

{\it Dechanneling length or Occupation length}: 
A simple estimate of the length over which particles dechannel due to multiple 
scattering is given by setting the rms multiple scattering angle equal to the 
Lindhard critical angle yields \cite{Baier}
\beq
L_{dechan} = \frac{\alp}{\pi}\left( \frac{U_0 E_e}{(m_e c^2)^2}\right) L_r
\label{eq: Ldechan}
\eeq
where $U_0$ is the depth of the atomic potential, $E_e$ is the particle energy.
This is based on a strictly classical approach
and predicts that the dechanneling length increases linearly with
energy. Measurements however have shown that the dechanneling lengths for 
electrons with energies in the tens of MeV are higher than the above
simple estimate and do not scale linearly with energy \cite{Kephart}. 

Quantum mechanically, a similar idea is expressed by the concept of an 
occupation length $L_{occ}$ which is the 
length over which the initial probability in a quantum state $n$ falls by a 
factor $1/e$, i.e.
\beq
P_n(z) = P_n(0) \exp[-\frac{z}{L_{occ}}]
\eeq
This occupation length depends on the states involved in the transition, the
plane of channeling and on the beam energy. It was measured in a few experiments, 
e.g. Refs.~\cite{Kephart, Nething}
with beam energies in the 5 - 54 MeV range. Some measured values for the (1 $\rarw$ 0)
transition in the (110) plane are shown in Table \ref{table: lengths}.

{\it Photon formation length}:
In the simplest version, this length represents the length scale over which the
photon ``shakes free'' from the electron after formation and separates from it
by a reduced wavelength $\lm/(2\pi)$ \cite{Uggerhoj}. It is given by 
\beq
L_f = \frac{2\gm^2 c}{\om}
\eeq
where $\om$ is the photon frequency. Clearly the crystal thickness should be
larger than this formation length for a significant photon yield.

{\it Photon absorption length}: 
The photon absorption length within a material is given by
\beq
\frac{1}{L_a} = \frac{N_A \sg_T}{A}
\eeq
where $N_A$ is the atomic density, $A$ is the atomic number and $\sg_T$ is the
total cross-section of all processes that lead to photon absorption during its
passage through the material. These include the photo-electric effect, 
Compton scattering, and also pair production for photon energies sufficiently above 
1 MeV. The scattering cross-sections for these processes are well known and the 
absorption lengths at different photon energies can be obtained from tables
maintained by NIST \cite{NIST_page}.

Table \ref{table: lengths} shows the values of these length scales for some
representative electron and photon energies. 
\begin{table}
\caption{Different length scales at different electron and photon energies.
$\dag$: the dechanneling length is found from Eq. \ref{eq: Ldechan} with a
depth $U_0=23.8$ eV for the (110) plane in diamond,
$\ddag$: the values for the occupation lengths are quoted for the (110) plane
and taken from Ref.~\cite{Kephart}.}
{\btable{|c|c|c|c|c|} \hline
 \multicolumn{5}{|c|}{Photon lengths} \\
 & $E_{\gm}$ [keV] & Length  &  $E_{\gm}$ [keV]   &  Length     \\ \hline
Formation length $L_f$ with 50 MeV e$^-$  & 10   & 0.38[$\mu$m]  &  80  & 0.047[$\mu$m] \\ 
Photon absorption length $L_a$ & 10  & 1.26 [mm]  &  80 & 17.7  [mm] \\
\hline
 \multicolumn{5}{|c|}{Electron lengths} \\ 
   & $E_e$ [MeV] & Length & $E_e$  [MeV] & Length \\ \hline
Electron radiation length $L_r$ & 20  & 16.7 [cm] & 50 & 14.9 [cm] \\
Dechanneling length $L_{dechan}\dag$  & 20  & 0.71 [$\mu$m] & 50  & 1.58 [$\mu$m] \\
Occupation length $L_{occ}\ddag$ (1 $\rarw$ 0)   &  17 & 20  [$\mu$m] &  54 
 &  36[$\mu$m]  \\
\hline
\etable \label{table: lengths} }
\end{table}
The crystal thickness $d$  should be large enough for enough photons to be emitted
from the particle, so $d > L_f$ but small enough that most photons do not get 
absorbed within the crystal, i.e. $d < L_a$. Since
in all cases of interest $d \ll L_r$, the radiation length, the electron will lose
very little of its energy through its passage through the crystal. We also note 
that the classical dechanneling length found from the simple estimate in 
Eq (\ref{eq: Ldechan}) significantly underestimates the occupation
length found from measurements at nearby energies.

\section{Simulations of ELBE experiments}

 We used a Mathematica notebook developed for modeling the channeling
radiation experiments at the ELBE facility \cite{Azadegan_ComPhys}. 
This notebook (called PCR) or some version of it was used to model the 
ELBE experiments and results in Ref.~\cite{Azadegan_thesis} showed the line 
widths were about half the measured values and photon yields were about a factor 
of two higher than the experimental 
results. However tests with the notebook available from the source 
\cite{Azadegan_ComPhys} showed much greater discrepancies with the ELBE
experimental results. We therefore  corrected and added features to the
notebook, the more significant changes are listed here in order of
importance: 
\benu
\item Used the set of equations (\ref{eq: Pn_mod_2}) to model the effects of 
enhanced
dechanneling  due to multiple scattering and other scattering phenomena
from the bound and quasi-free states.
\item Corrected the imaginary part of the potential as described in 
Section \ref{subsec: NonRad}.
This also included correcting the matrix elements of
the imaginary part of the potential. These matrix elements now better
obey the approximate selection rule that the non-radiative transitions
occur primarily between states of the same parity. The Fourier coefficients
of the real and imaginary parts of the potential are now also calculated in the 
notebook. 
\item Included the effects of a finite beam divergence.
\item Included the contributions to the linewidth from Bloch-wave broadening
and multiple scattering to the total line width. The notebook in 
Ref.~\cite{Azadegan_ComPhys} contained only the contribution from the coherence length.
\item Corrected the line shape in the intensity spectrum calculation
\item Included photon self-absorption within the crystal.
\eenu

Further improvements could be made to the physics model. These include:
\bit
\item The effect of inelastic electron scattering off the valence and bound 
electrons on the transition matrix elements $W_{mn}$ needs to be included.
The importance of electron scattering to the linewidth was discussed in 
Ref.~\cite{Genz_96} and will be discussed in Section \ref{subsec: ELBE_ELWY}
below. 
\item The rms multiple scattering angle while channeling in a crystal is obtained 
from that in
an amorphous medium by a scaling factor, based on limited experimental
data. This could be replaced by a calculation of the multiple scattering angle
during channeling from first principles. 
\eit

Table \ref{table: ELBE_param} shows the main parameters of the ELBE facility
which we used in the simulations reported here. 
\begin{table}[h]
\caption{Main parameters of the electron beam in ELBE, Ref.~\cite{Wagner}}
{\btable{|c|c|} \hline
Crystal thickness[$\mu$m] & 42.5, 168, 500   \\
Beam energy [MeV] & 14.6, 17, 30, 34   \\
Uncertainty in beam energy [MeV] & 0.2 \\
Transverse norm. emittance [mm-mrad] & 3.     \\
Beam divergence, both planes [mrad] & 0.1  \\
Relative energy spread & $1.3\times 10^{-3}$  \\
Beam size at crystal [mm] &  $\sim 1$    \\
Average beam current [nA] & $\ge 100 $ \\
\hline
\etable \label{table: ELBE_param}}
\end{table}

First we discuss the coherence length calculation.
Since the corrected imaginary part of the potential is weaker that that used
in the previous calculations for ELBE, the coherence length found here is 
longer and hence the linewidth from this effect is smaller than that reported in 
Ref.~\cite{Azadegan_PRB}. 
We note that this coherence length includes only the effect of 
inelastic thermal scattering and does not include the effects of inelastic
electron scattering. 
Table \ref{table: Lcoh_ELBE} lists the coherence length from thermal scattering
at the ELBE energies for some low order transitions
\begin{table}
\caption{Coherence lengths ($\mu$m) for the low order transitions}
{\btable{|c|c|c|c|} \hline
Beam energy [MeV] &  $L_{coh}$ $1\rarw 0$ [$\mu$m] & $L_{coh}$ $2\rarw 1$ [$\mu$m]
 & $L_{coh}$  $3\rarw 2$ [$\mu$m] \\ \hline
14.6 & 1.87 & 4.08 & 5.98   \\
17.0 & 1.79 & 3.75 & 5.95  \\
25.0 & 1.62 & 3.01 & 5.24 \\
30.0 & 1.55 & 2.74 & 4.70  \\
\hline
\etable \label{table: Lcoh_ELBE}}
\end{table}
The coherence length for the $1\rarw 0$ transition extracted from measured
data \cite{Azadegan_PRB} was about 0.65 $\mu$m, more than a factor of two 
smaller than the
calculated values. This discrepancy is most
likely due to neglecting the inelastic electron-plasmon and
electron-core electrons from the atom scattering.

\subsection{Population dynamics}\label{subsec: population}

The populations in the  states change as the electron moves through the 
crystal.
In our simulations we have modeled dechanneling due to all effects in a heuristic 
fashion using the
model with the free parameter $n_f$ in Eqs. (\ref{eq: Pn_mod_1}) and 
(\ref{eq: Pn_mod_2}). Let $n_B$ denote the index of the highest bound state
at a given energy; $n_B=3$ at 14.6 MeV and $n_B=4$ at 30 MeV with the ground state
index $n=0$. 
For each energy and crystal thickness, the appropriate values of $n_f$ were
determined by comparing with the measured photon yield, to be discussed below in 
Section \ref{subsec: ELBE_ELWY}.  Lower values of
$n_f$ imply dechanneling from more free states and hence lower photon yields.

\begin{figure}[h]
\centering
\includegraphics[scale=0.38]{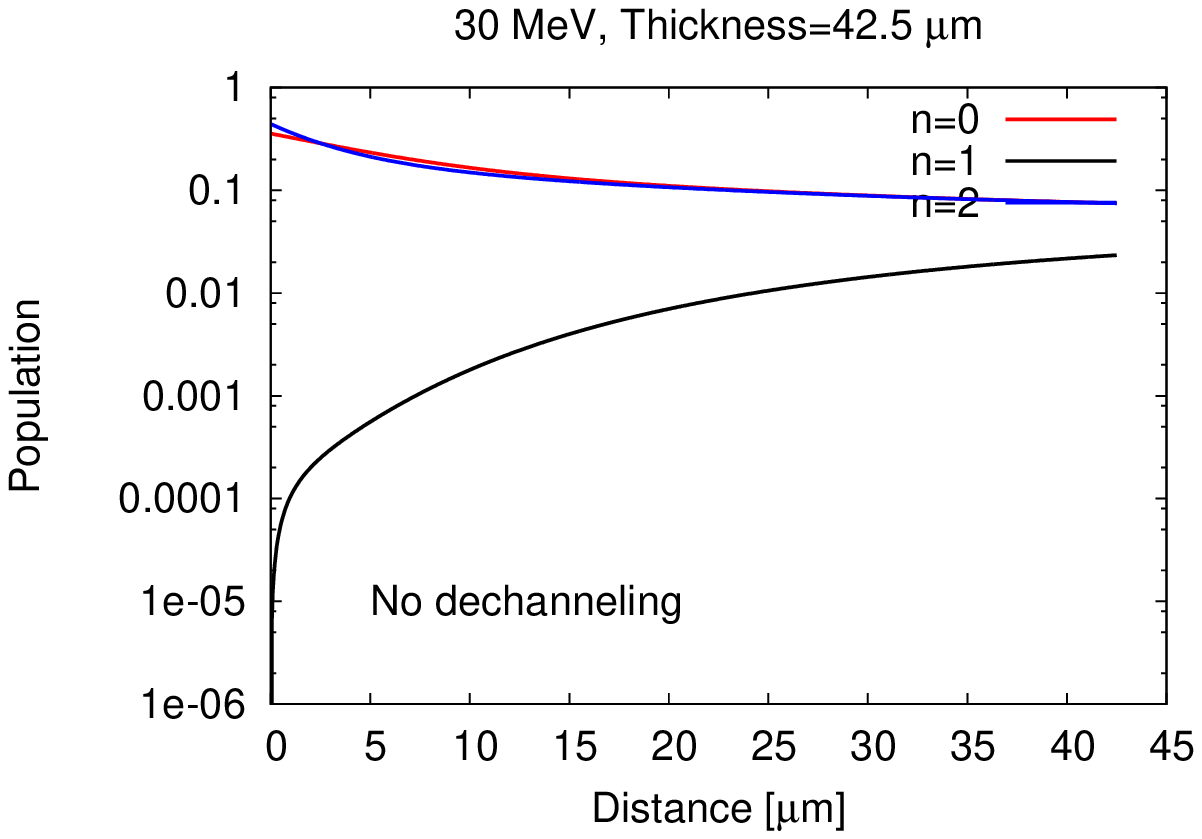}
\includegraphics[scale=0.38]{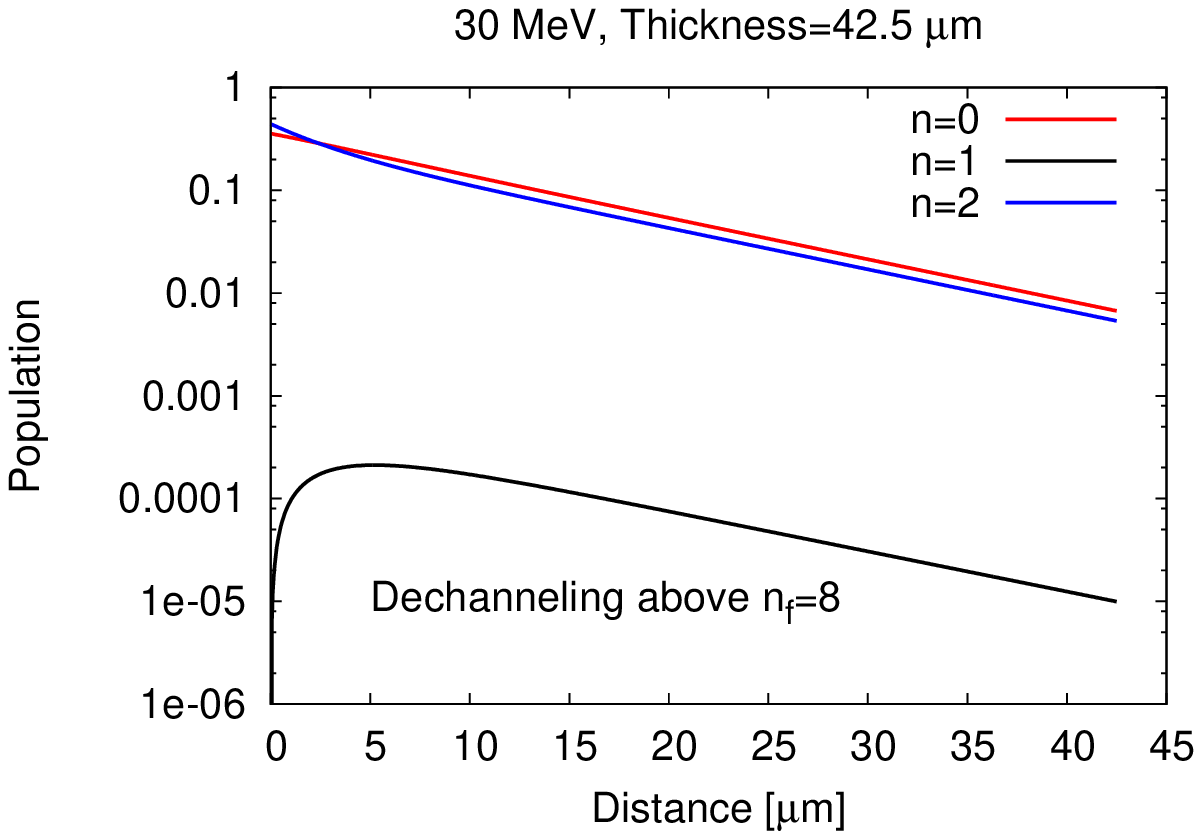}
\includegraphics[scale=0.38]{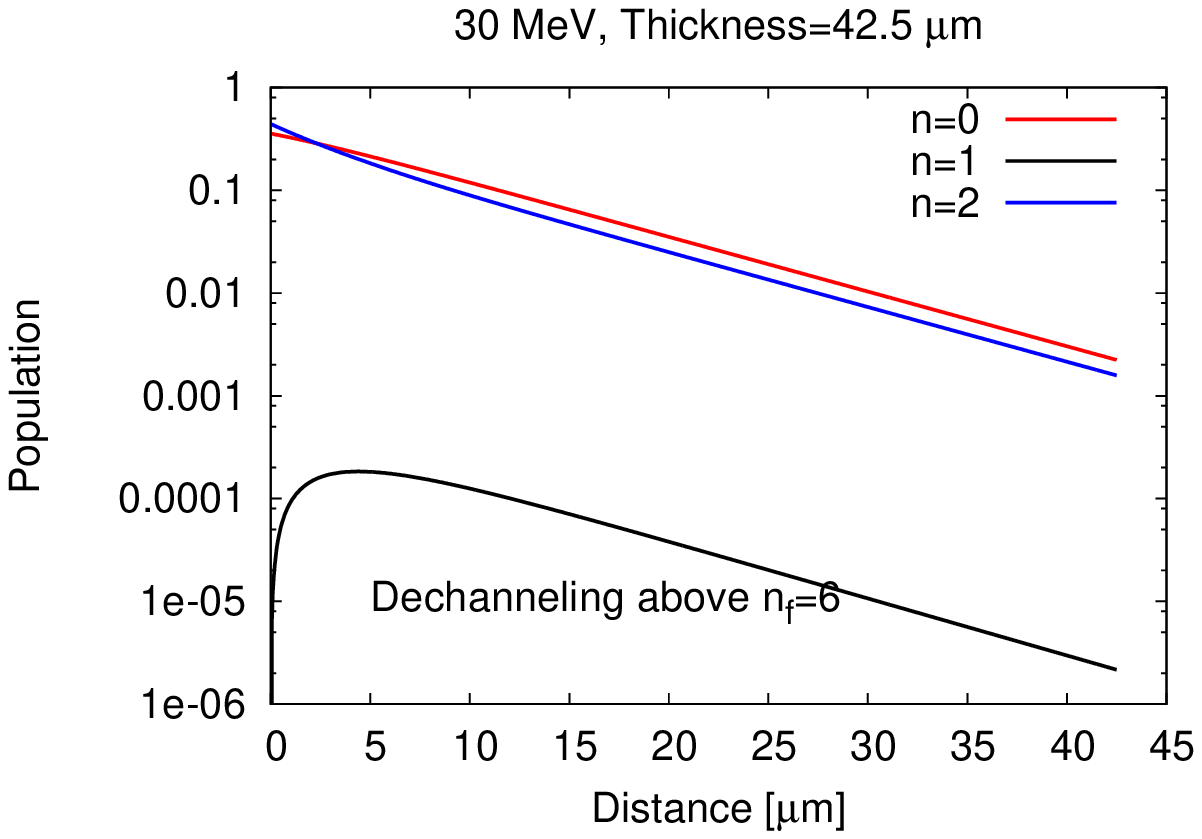}
\caption{Population of electrons (energy=30 MeV) vs distance into crystal 
(thickness=42.5 $\mu$m) for 3 cases. 
Left: no dechanneling, Middle: $n_f=8=n_B+4$, Right: $n_f=6 =n_B+2$.}
\label{fig: popz_ELBE_5}
\centering
\includegraphics[scale=0.38]{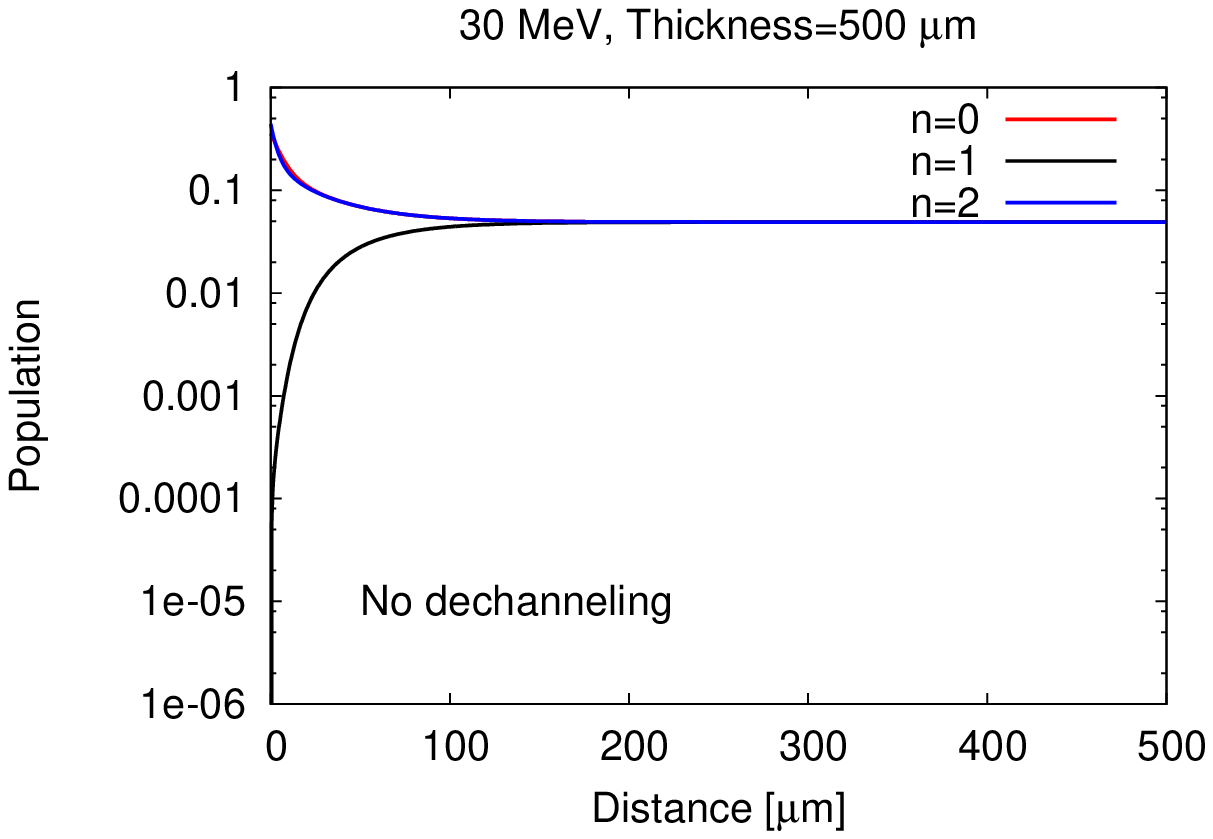}
\includegraphics[scale=0.38]{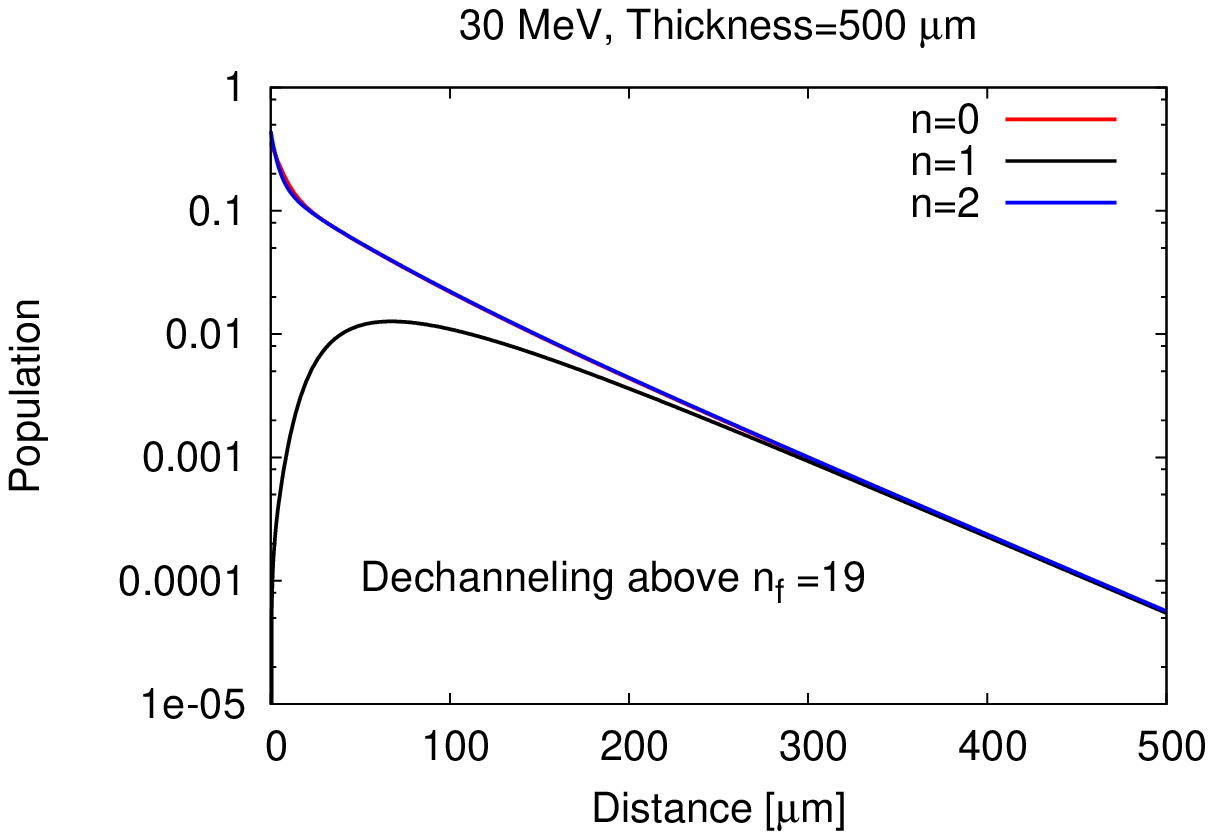}
\includegraphics[scale=0.38]{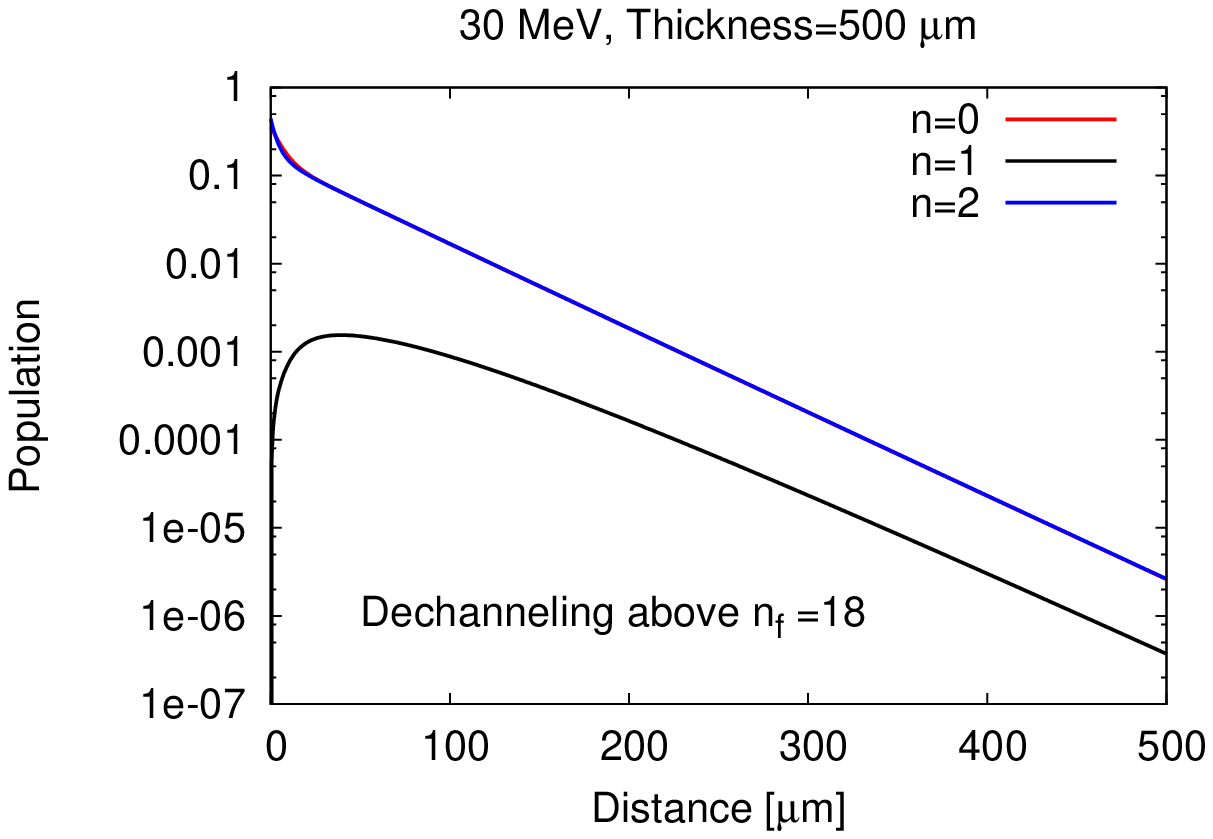}
\caption{Population vs distance into crystal (thickness=500 $\mu$m), electron 
energy=30 MeV for 3 cases of dechanneling. 
Left: no dechanneling, Middle: $n_f=19$, Right: $n_f=18$.}
\label{fig: popz_ELBE_10}
\end{figure}

Here we discuss the probabilities $P_n(z)$ found from the numerical solutions
of Eqs. (\ref{eq: Pn_mod_1}), (\ref{eq: Pn_mod_2}). Figures \ref{fig: popz_ELBE_5}
and \ref{fig: popz_ELBE_10} show the populations as a function of the distance 
into the crystal for thicknesses of 42.5 $\mu$m and 500 $\mu$m respectively at the
beam energy of 30 MeV. Only the three lowest bound states are shown in each case.
The appropriate values of $n_f$ change with the thickness. The left plot in both
figures show the populations without enhanced dechanneling ($n_f=21$). 
Without the additional 
dechanneling, the populations in the three states equalize and reach equilibrium
at around 200 $\mu$m, as seen in Fig. \ref{fig: popz_ELBE_10}.
This length is relatively insensitive to the energy, being about the same at all 
energies modeled. Without dechanneling from multiple scattering, the photon yield
was significantly higher than the experimental value. 

Two values of $n_f$ were chosen such that the measured yield lay in between the 
calculated photon yields with these values of $n_f$. At 42.5 $\mu$m thickness,
these values at 30 MeV were $n_f = (6, 5)$ while at 500 $\mu$m, these were 
$n_f= (19, 18)$. The middle plot in Figures \ref{fig: popz_ELBE_5} and 
\ref{fig: popz_ELBE_10} shows 
the populations with the higher values of
$n_f$ in each case. As expected, the populations in these states do not reach
equilibrium but continue to decrease with distance. At 500 microns, the three
states have nearly equal populations and fall at the same rate implying that
the occupation lengths are about the same in these states. 
The right plot in both figures shows the populations when $n_f$ is set so that
the calculated yield is a slight under-estimate of the measured yield. For both
thicknesses, we find that the populations in the $n=1$ state increase by
several orders of magnitude before reaching a maximum and 
falling at the same rate as in the even bound states.
We find that if $n_f$ is so low that the initial increase in the $n=1$ state
is less than an order of magnitude, the 
dechanneling is too strong and the photon yield is much lower 
than the measured value. The dependence of $P_n(z)$
for different values of $n_f$ is similar at other energies. 

\begin{figure}[h]
\centering
\includegraphics[scale=0.55]{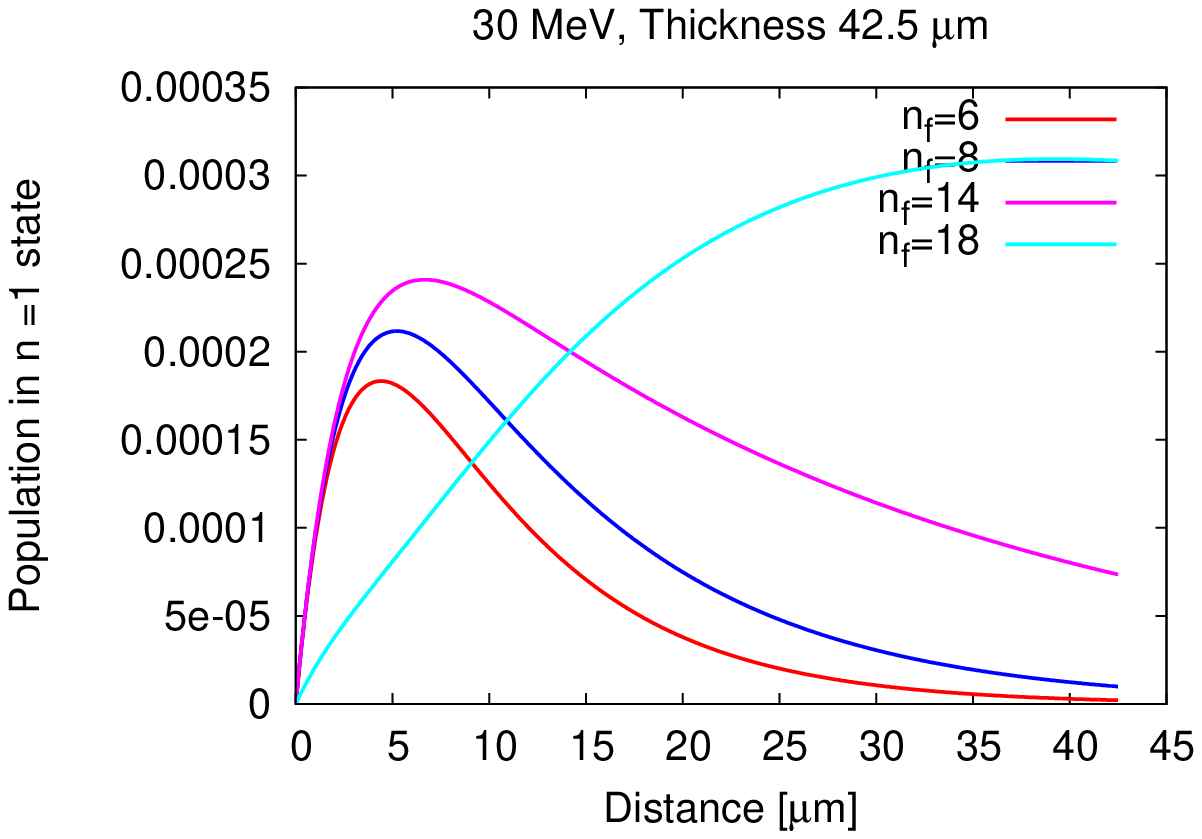}
\includegraphics[scale=0.55]{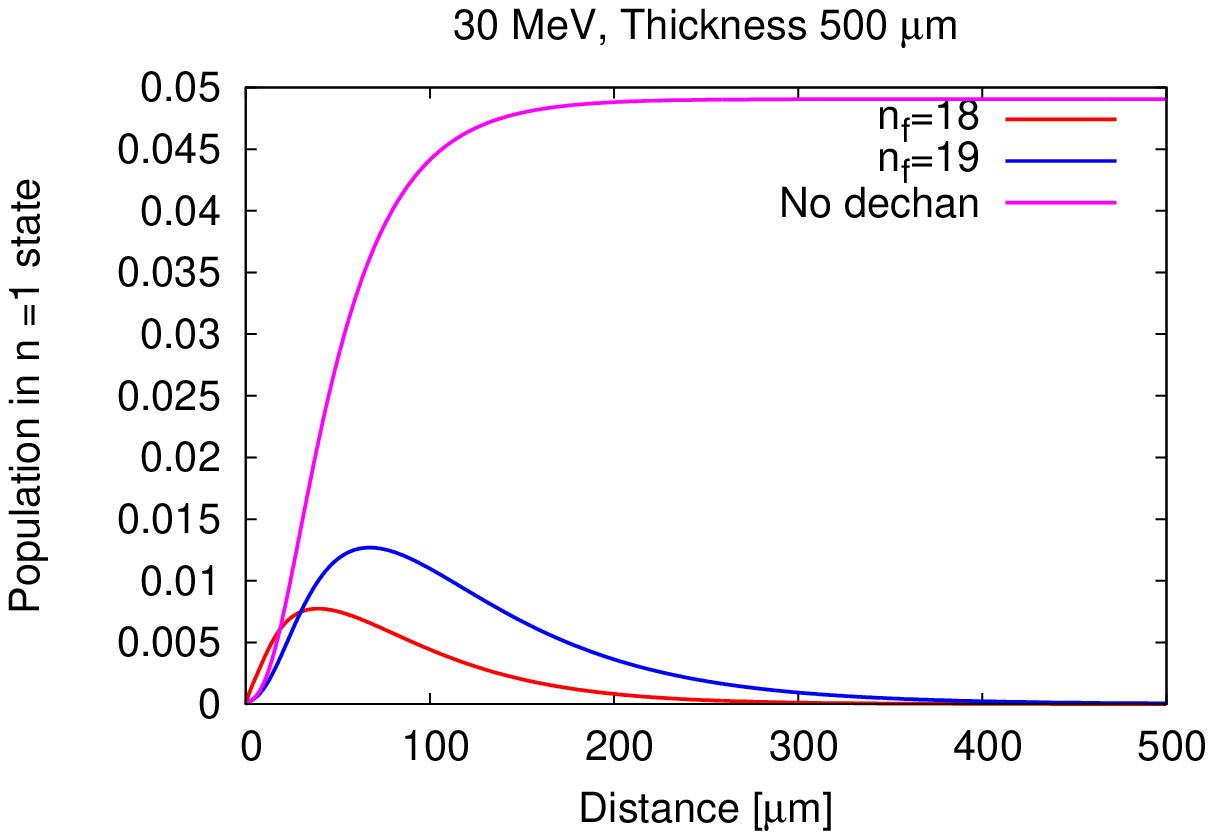}
\caption{Population in state $n=1$ at energy = 30 MeV. Left:
thickness = 42.5 $\mu$m), Right: 500 $\mu$m. 
Populations are shown for different values of the lowest
free state $n_f$ from which dechanneling occurs. 
The smallest value of $n_f$ shown in each plot is the
one for which simulated yields best match the experimental yields.
At 42.5 $\mu$m, $P_1(z;n_f=18)$ is scaled down by 0.005, while
at 500 $\mu$m $P_1(z;n_f=18)$ is scaled up by 5, in order to
show all populations on a linear scale. }
\label{fig: pop_n=1_3-4}
\end{figure}
Fits to the population in the lowest order even states $n=0,2$
show good fits to an exponential form 
$P_n(z) = P_n(0)\exp[-z/L_{occ}]$, especially at the lower energies. At
30 MeV, a better fit is obtained with a sum of two exponentials of the
form $P_n(z) = P_n(0)[\exp[-z/L_{n1}] + a\exp[-z/L_{n2}]]$, where
 $L_{n2} \gg L_{n1}$. However the weight of the second term is small, 
typically $a \sim 0.3$. \newline
Since the intensity of the $1\rarw 0$ transition is determined by the
population in the $n=1$ state, we consider it in a little more detail.
Figure  \ref{fig: pop_n=1_3-4} shows the populations in this state
at 30 MeV and different thicknesses at a few chosen values of the 
parameter $n_f$. For both thicknesses we observe that as $n_f$ decreases,
the distance at which the population reaches a maximum decreases and also
decays at a faster rate, i.e. with a shorter occupation length. 
The behavior shown can be modeled by a functional form
\beq
 P_1(z) =  (\frac{z}{L_0})^q \exp[-\frac{z}{L_{occ}}]
\label{eq: P1_z}
\eeq
Here $L_0$ is a length parameter determined by the maximum of the population
while the distance at which the population is maximum is given by 
$z_m = q L_{occ}$. 
Table \ref{table: occ.length} shows fitted values of $L_{occ}$ and $q$
at different beam energies and different thicknesses. The fits for the 
occupation length $L_{occ}$ in the bound states $n=0, 2$ yield very similar
values to those shown in this table. 
\begin{table}
\caption{Values of the occupation length and the power law exponent $q$ 
for two energies and thicknesses.}
{\btable{|c|c|c|c|c|} \hline
Energy & Thickness & $n_f$ & Occ. length $L_{occ}$ [$\mu$m] & $q$ \\ \hline
14.6 & 42.5 & 4 &  8.6 & 0.93 \\
14.6 & 500 & 16 & 39.5 & 0.19 \\
30.0 & 42.5 & 6 & 8.2 & 0.54 \\
30.0 & 500 & 18 & 52.7 & 0.75 \\
\hline
\etable \label{table: occ.length}}
\end{table}
We observe that the occupation length changes relatively little with energy but
depends strongly on the thickness. This is one indication that the rechanneling
probability which increases with thickness has a strong impact on the
population dynamics. Rechanneling occurs when an electron in a dechanneled
free state enters a bound state by losing transverse energy due to a number of
processes including multiple scattering. It is possible that the rechanneling
probability and the occupation length saturate for sufficiently thick
crystals. Nevertheless from the results in Table 
\ref{table: occ.length} we can conclude that occupation lengths 
cannot be considered in isolation from rechanneling and crystal thickness and
that classical expressions for the dechanneling length such as in Eq.
(\ref{eq: Ldechan}) may be invalid in the quantum regime. 

Equation (\ref{eq: P1_z}) can be used to estimate the crystal thickness at 
which the intensity of the $1 \rarw 0$ transition will saturate. The 
intensity for a crystal of thickness $d$ relative to an infinitesimally thin 
crystal in the limit that the photon absorption length $L_a$ is long compared 
to the crystal thickness is proportional to the integral of $P_1(z)$,
\beq
I(d) \propto \int_0^d P_1(z) dz = L_{occ}(\frac{L_{occ}}{L_0})^q
\left[ \Gm(1+q) - \Gm(1+q,\frac{d}{L_{occ}}) \right]
\eeq
where $\Gm$ is the gamma function. 
\begin{figure}[h]
\centering
\includegraphics[scale=0.8]{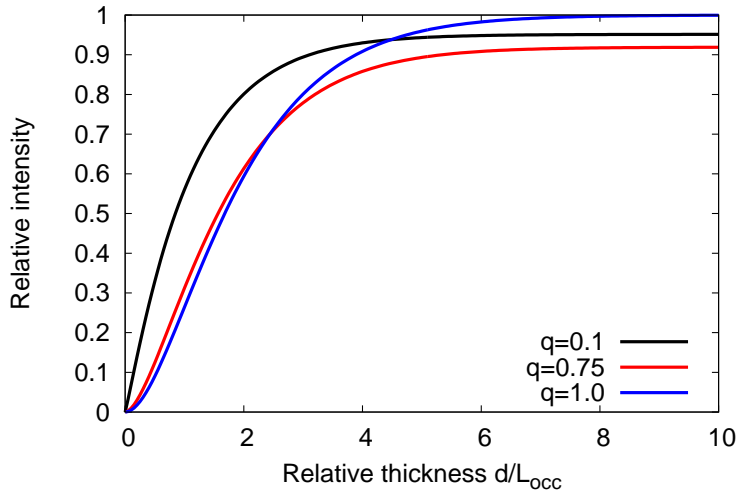}
\caption{Relative intensity for the 1$\rarw 0$ transition as a function of the
crystal thickness $d$ relative to the occupation length $L_{occ}$ and 
different values of the power law parameter $q$ in Eq.(\ref{eq: P1_z}).}
\label{fig: rel_intens_1-0}
\end{figure}
Figure \ref{fig: rel_intens_1-0} shows the relative intensity as a function of
the relative crystal thickness $d/L_{occ}$ for three values of the power law
exponent $q$. 
Figure \ref{fig: rel_intens_1-0} shows that the intensity of the $1\rarw 0$ 
transition saturates within a thickness of $d=7 L_{occ}$ for the range of $q$
values considered. 
 This suggest that when $L_{occ} \simeq (40-50)$ $\mu$m as seen in 
Table \ref{table: occ.length}, crystal
thickness of $\sim 350$ $\mu$m may suffice to optimize the channeled
fraction and the intensity of the $1\rarw 0$ transition.

\subsection{X-ray energies, line widths and photon yields} \label{subsec: ELBE_ELWY} \label{sect: ELBE_EWY}

Here we discuss the main aspects of the X-ray photon spectrum for the ELBE 
parameters.  We have assumed here and in subsequent calculations that 
variations in the incidence angle from zero are small compared to the
beam divergence. Table \ref{table: ELBE_En_Wid}
shows the energies of the 1$\rarw$ 0 transition for different beam energies and different crystal thicknesses. 
\begin{table}
\caption{Results for X-ray energies and line widths using a diamond crystal in
 the (110) plane and the $1\rarw 0$ transition. Experimental values from 
Tables IV and V in Ref.~\cite{Azadegan_PRB}.}
{ \btable{|c|c|c|c|c|c|} \hline
$e^-$ Energy  & Thickness & \multicolumn{2}{|c|}{Energy[keV]} & 
\multicolumn{2}{|c|}{Linewidth [keV]} 
 \\
\mbox{} [MeV]  & $\mu$m & $E_{exp}$ & $ \lan E_{sim} \ran $ 
& $\tilde{\Gm}_{exp}$ & $\Gm_{sim}$ \\ \hline
14.6 & 42.5 & 16.58 & 16.35  & 1.43 & 0.74  \\
     & 168 &  16.99 & 16.01 & 1.74 & 1.09   \\
     & 500 &  16.47 & 15.63   & 2.15 & 1.49  \\
\hline
17 & 42.5 & 21.72 & 21.41 & 1.94 & 0.92  \\
   & 168 & 22.37 & 20.97   & 2.35 & 1.40  \\
   & 500 & 21.38 & 20.48  & 2.73  & 1.92 \\
\hline
25 & 42.5 & -  & 42.02  & -   &  1.78 \\ 
\hline
30 & 42.5 & 56.19 & 57.58  & 5.85 & 2.46 \\
   & 168 & 56.22 & 56.44  & 6.09 & 3.72 \\
   & 500 & 55.06  & 55.13  & 11.96   & 5.09   \\
\hline
\etable \label{table: ELBE_En_Wid} }
\end{table}
In order to mimic the experimentally observed dependence of the X-ray energy on the thickness, we have used
$\lan E_{sim} \ran$,  the average value of the peak due to
Doppler shift from multiple scattering, given in Eq(\ref{eq: Eav_MS}). Due to 
the greater multiple scattering in thicker crystals, the value of 
$\lan E_{sim} \ran$ decreases with thickness. This trend is also observed in 
the experimental values when going
from 42.5 to 500 $\mu$m at all energies but for 168 $\mu$m only at one energy. 
In most cases, the simulated value agrees to within 6\% which is well within 
the error bars on the measurements. 

In comparing the linewidths, we have defined  the quantity $\tilde{\Gm}_{exp}$
 as the experimental line width but with the detector energy resolution 
removed via a quadrature, i.e.
$ \tilde{\Gm}_{exp} = \sqrt{\Gm_{exp}^2 - \Gm_{det}^2} $
where $\Gm_{exp}$ is the measured linewidth. Table \ref{table: ELBE_En_Wid} shows that the  simulated values are 
consistently smaller than the measured values, in some cases by  more than half. The most likely reason for this 
under-estimate is the neglect of scattering off 
the atomic electrons. 
 Genz et al \cite{Genz_96} had  concluded from their measurements that 
electronic scattering is not negligible in its contribution to the linewidth. In
principle, the imaginary part of the  potential for electron-electron scattering
$V_{el}^I$ could also cause non-radiative transitions and should be added to the
potential for phonon scattering. 
However since the momentum transfers involved in electronic scattering are small,
the transition rates $\lan m | V_{el}^I | n\ran$ for $m \neq n$ are small and
therefore  the transition rates to neighboring  and more distant energy bands will be small. Thus their contributions to the
population dynamics can most likely be ignored. 
However the linewidths involve
the expectation values of the potential in the states involved in the transition, 
$\lan m | V_{el}^I | m\ran$ etc and these can be comparable to the values for
thermal scattering.

Table \ref{table: ELBE_yields} shows the photon yields for different beam
energies and crystal thicknesses. In each case the yields are shown for 
three values of $n_f$; one corresponding to no (enhanced) dechanneling, and the other
two for which the simulated yields are closest to the experimental yield. 
$n_B$ is the index of the highest bound state which changes with the energy. 
\begin{table}
\caption{Results for diamond crystal with the (110) planes and the
$1\rarw 0$ transition. Experimental values from Tables 2.2 and 2.6 in 
Ref.~\cite{Azadegan_thesis}.
For the simulated yields, no (enhanced) dechanneling was considered in one case
while the other two had this dechanneling included in the model with different
values of $n_f$ relative to the index $n_B$ for the highest bound state. }
{\begin{tabular}{|c|c|c|c|c|c|} \hline
$e^-$ Energy  & Thickness & 
\multicolumn{4}{|c|}{Yield $dN/d\Om$ [phot/e-/sr]} \\
\mbox{} [MeV]  & $\mu$m & Exp. yield & \multicolumn{3}{|c|}{Sim yield} \\
 &   &   & No dechan. $Y_{sim}$ & $n_f$ $|$ $Y_{sim}$ & $n_f$ $|$ $Y_{sim}$ \\
\hline
14.6 & 42.5 & 0.048  & 0.129 & $n_B+3$ $|$ 0.053 & $n_B+1$  $|$ 0.044  \\
     & 168 &   0.090   & 0.36 & $n_B+9$  $|$ 0.11  & $n_B+7$ $|$ 0.089 \\
     & 500 &   0.149   & 0.89 &  $n_B+15$ $|$ 0.18 & $n_B+14$ $|$ 0.16  \\
\hline
17 & 42.5 & 0.059   & 0.18 &  $n_B+3$  $|$ 0.069  & $n_B+1$ $|$ 0.057 \\
   & 168  &  0.13   & 0.52 & $n_B+9$ $|$ 0.15 & $n_B+7$  $|$ 0.12 \\
   & 500  &  0.30 & 1.31 &  $n_B+16$ $|$ 0.34 &  $n_B+15$ $|$ 0.26     \\
\hline
25 & 42.5 & 0.159 & 0.45  & $n_B+3$ $|$ 0.14 & $n_B+1$  $|$  0.11 \\ 
\hline
30 & 42.5 &   0.229 & 0.68  & $n_B+3$ $|$ 0.24 & $n_B+1$  $|$ 0.18 \\
   & 168  &   0.52  & 1.64  & $n_B+9$ $|$ 0.54  & $n_B+7$ $|$ 0.43 \\
   & 500  &   1.012  & 5.23 &  $n_B+15$ $|$ 1.33  &  $n_B+14$ $|$ 0.95 \\
\hline
\end{tabular} \label{table: ELBE_yields}}
\end{table}
From the results shown in Table \ref{table: ELBE_yields} we observe first that
without dechanneling, the photon yields in the model are significantly higher
than experimental values in all cases and at the same energy, the difference 
increases with crystal thickness. This is a clear indication that dechanneling
effects need to be included in the model. The last two columns show the
simulated yields when these are included. We observe that the lowest free state 
$n_f$ relative to the highest bound state $n_B$ depends almost entirely on
the crystal thickness. Thus with 42.5 $\mu$m, the experimental yield is bounded by
the yield in the states
$(n_B+3, n_B+1)$ at all energies, with 168 $\mu$m, the relevant states are
$(n_B+9, n_B+7)$ again at all energies while with 500 $\mu$m, the relevant
states are $(n_B+15, n_B+14)$ at 14.6 MeV and 30 MeV and $(n_B+16, n_B+15)$
at 17 MeV. 

The fact that these bounding states depend only on the thickness and not
on the energy is both
significant and useful. It shows a) that the energy dependence in the model is
reasonably accurate and b) the conjecture that the experimental yield is
obtained by including dechanneling effects which increase with
crystal thickness is most likely correct. It is useful because results
obtained with a given crystal thickness at a certain energy can be used to
predict the yields at other energies. We will use this feature in the next
section to estimate the photon yields with ASTA parameters. 

Another significant conclusion inferred from the results in Table
\ref{table: ELBE_yields} is that rechanneling is important,
especially for thicker crystals. We find that for a thickness of 42.5 $\mu$m,
the assumption that dechanneling occurs from bound and nearly all the free states 
is a good 
model. This follows from the observation that the theoretical yield with
$n_f = n_B+1$ or $n_f = n_B+3$ are the closest to the experimental yield at 
this thickness.
For thicker crystals, such low values of $n_f$ leads to yields much smaller than 
experimental values. 
The fact that $n_f$ in the model increases with thickness in order to
match the experimental yields shows that rechanneling
significantly affects the observed yield.
The relative absence of rechanneling in thin crystals would explain why only
the populations in the bound states and the lowest free states can be 
considered
to contribute to the radiation yield. For thicker crystals this is a likely a
wrong assumption that drastically reduces the yield. Instead the electrons 
which
are in the higher free states can also scatter back into the bound states and 
increase the yield by radiative emission. 

\begin{figure}[h]
\centering
\includegraphics[scale=1]{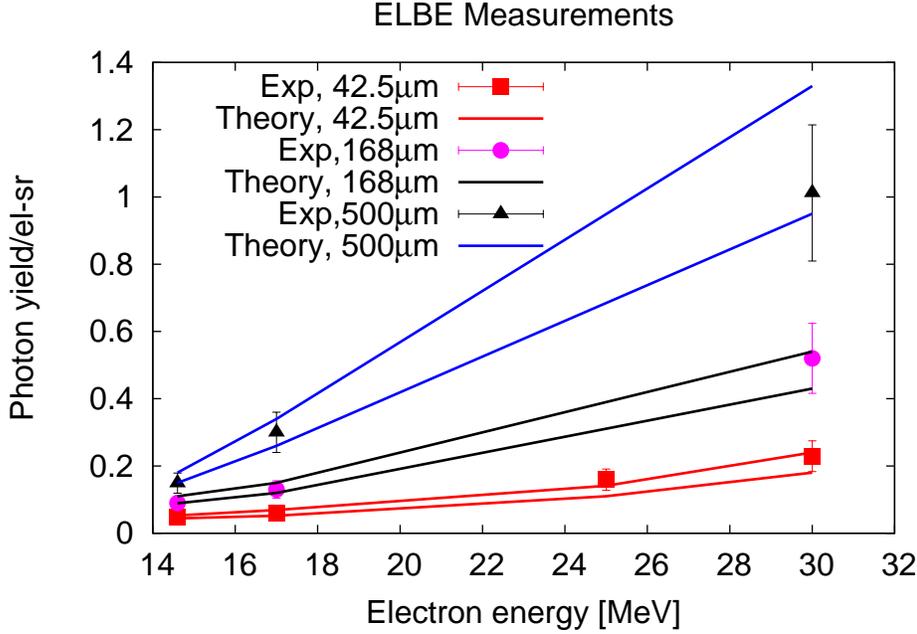}
\caption{Comparison of the experimental yield including 20\% error bars 
with the theoretical values found with the updated model.
At each thickness, the lower and upper solid lines correspond to the
smaller and larger values of $n_f$ in Table \ref{table: ELBE_yields}.
}
\label{fig: elbe_compare}
\end{figure}

Figure \ref{fig: elbe_compare} shows a comparison of the experimental yields
with the 20\% error bars quoted in Ref.~\cite{Azadegan_thesis} as a function of 
energy with the simulated photon yields from the two bounding states 
with dechanneling. 
We observe that the lower value of the simulated yield is within 15\% of the
experimental yield in most cases. This compares to nearly a factor of two 
difference in the earlier simulations \cite{Azadegan_thesis}. 
It is clear that the simulation model with the lower value of $n_f$ can be
used to predict the expected yield over this range of thicknesses. We will
use the population equations, Eqs. (\ref{eq: Pn_mod_1}), (\ref{eq: Pn_mod_2}),
with the lower $n_f$ to calculate the 
expected yield with ASTA parameters.

%\clearpage

\section{ASTA simulations}

In this section we apply the model to the ASTA photoinjector
and calculate the expected X-ray properties including the brilliance. 
The main parameters of ASTA are shown in Table \ref{table: ALBA_param}.
The major improvement over the ELBE facility is in the transverse
emittance of the electron beam. Recent developments have shown that
normalized emittances of less than 100 nm can be obtained with a
conventional laser photocathode by suitably reducing the laser spot size
\cite{Li_12}. Recent studies of field emission based cathodes using needle
like structures with tips of 5 nm radius of curvature have shown promising 
results \cite{Gabella}. Estimates show that the normalized emittances of 
the electron beam at the needle cathode can be as small as 1 nm. Simulations have 
shown
that this emittance is mostly preserved from the source to the crystal about
5m downstream. 
Here however we will assume a 
normalized emittance of 100 nm. Reductions in this emittance will 
increase the spectral brilliance. Diamond crystals cut parallel to the (110) 
plane are already available and these will be used for all the studies reported
here. 
\begin{table}
\caption{ASTA beam parameters at two different electron energies. A diamond
crystal will be used cut parallel to the (110) planes.}
{ \btable{|c|c|c|} \hline
Beam energy [MeV] & 20 &  50 \\
Bunch charge [pC] & 20 &  20 \\
Bunch frequency [MHz] & 3 &  3  \\
Average beam current [nA] & 300 & 300 \\
Transverse normalized emittance [nm] & $\le 100 $ & $\le 100 $ \\
Bunch length [mm] & $\le 1$  & $\le 1$ \\
Relative energy spread [\%]& $\le 1$  & $\le 1$   \\
Critical angle [mrad] & 1.54 &  0.98 \\
\hline
\etable \label{table: ALBA_param} }
\end{table}

\subsection{Potential and Populations}

\begin{figure}[h]
\centering
\includegraphics[scale=0.35]{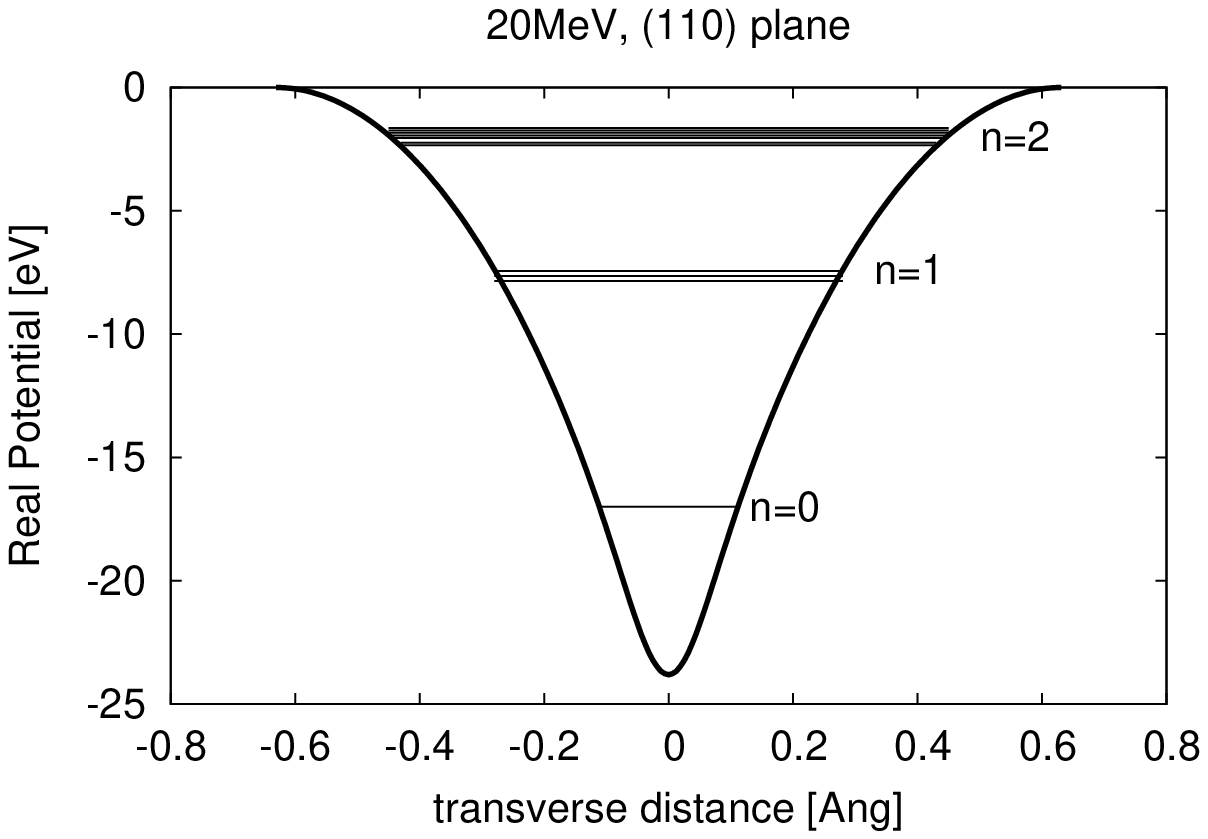}
\includegraphics[scale=0.35]{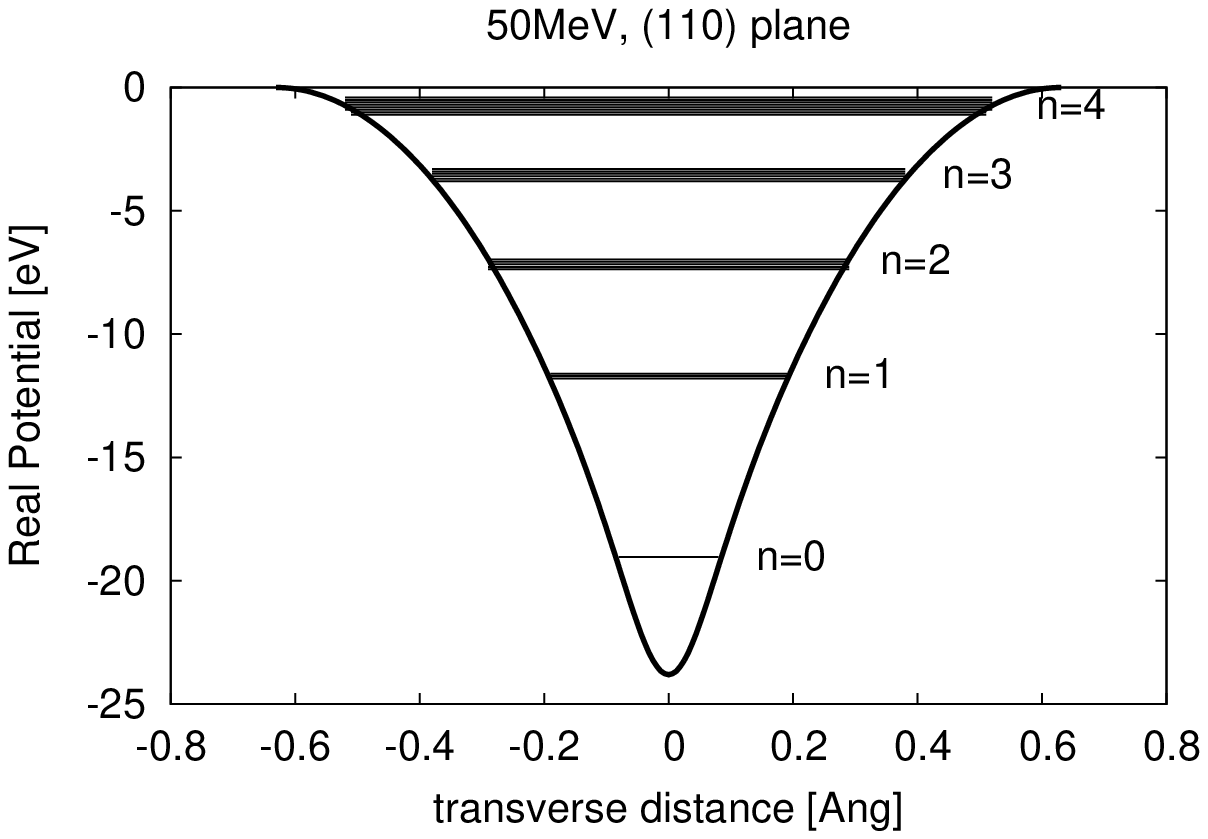}
\includegraphics[scale=0.35]{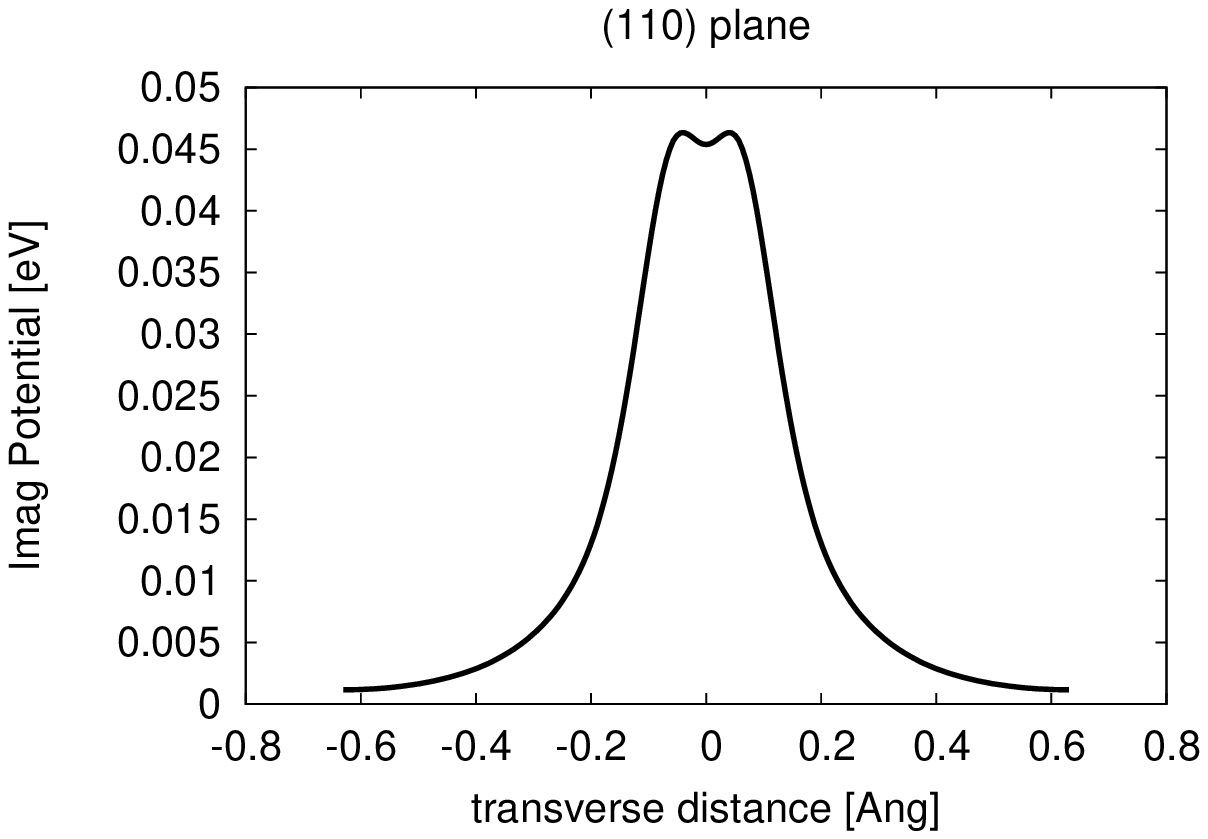}
\caption{Left: Real potential with bound state levels at 20 MeV; 
Middle: Real potential at 50 MeV, Right: Imaginary potential}
\label{fig: potentials_50}
\end{figure}

Figure \ref{fig: potentials_50} shows the real potential with the bound
states at 20 MeV and 50 MeV and the imaginary potential. These potentials
depend on the crystal lattice and the chosen planes while the number of bound
states (shown as bands in the two figures) increase with beam energy roughly
as $\gm^{1/2}$. The depth of the real potential for the (110) plane in 
diamond is about 23.8 eV while the height of the imaginary potential is about
0.045 eV. 

\begin{figure}[h]
\centering
\includegraphics[scale=0.5]{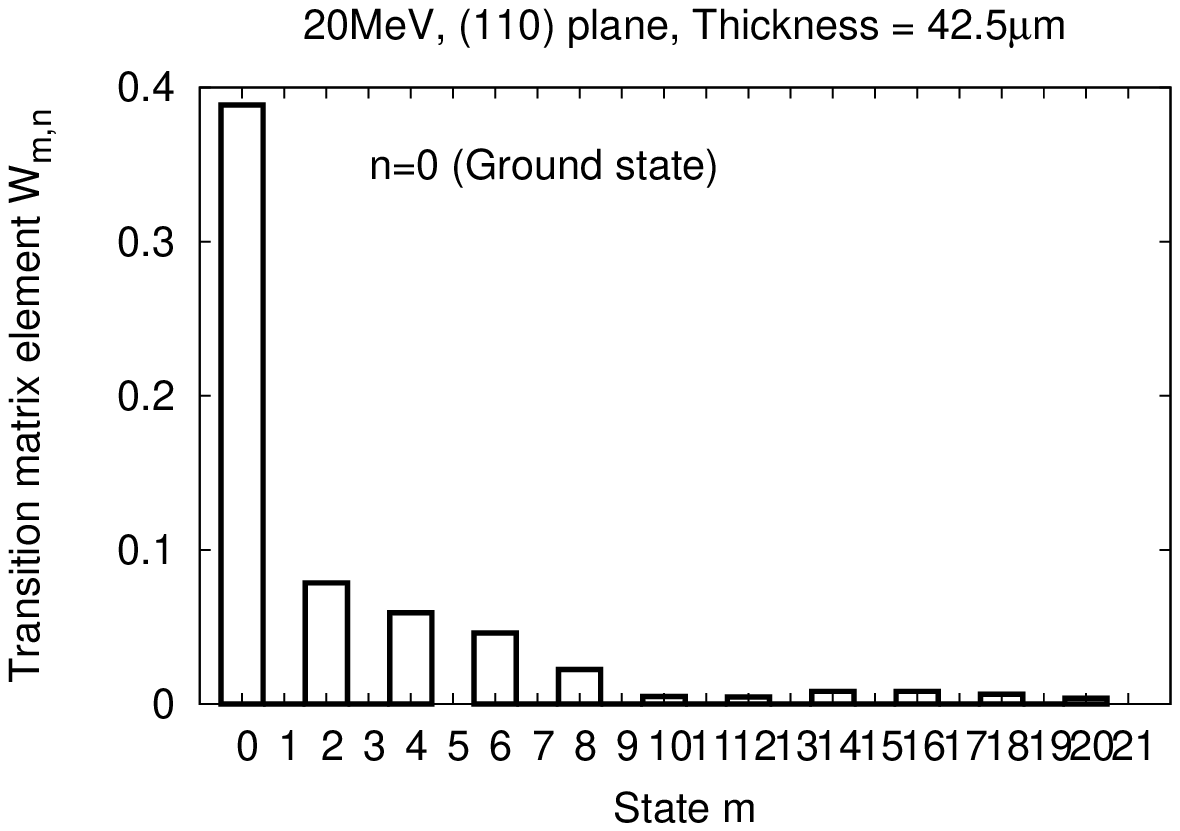}
\includegraphics[scale=0.5]{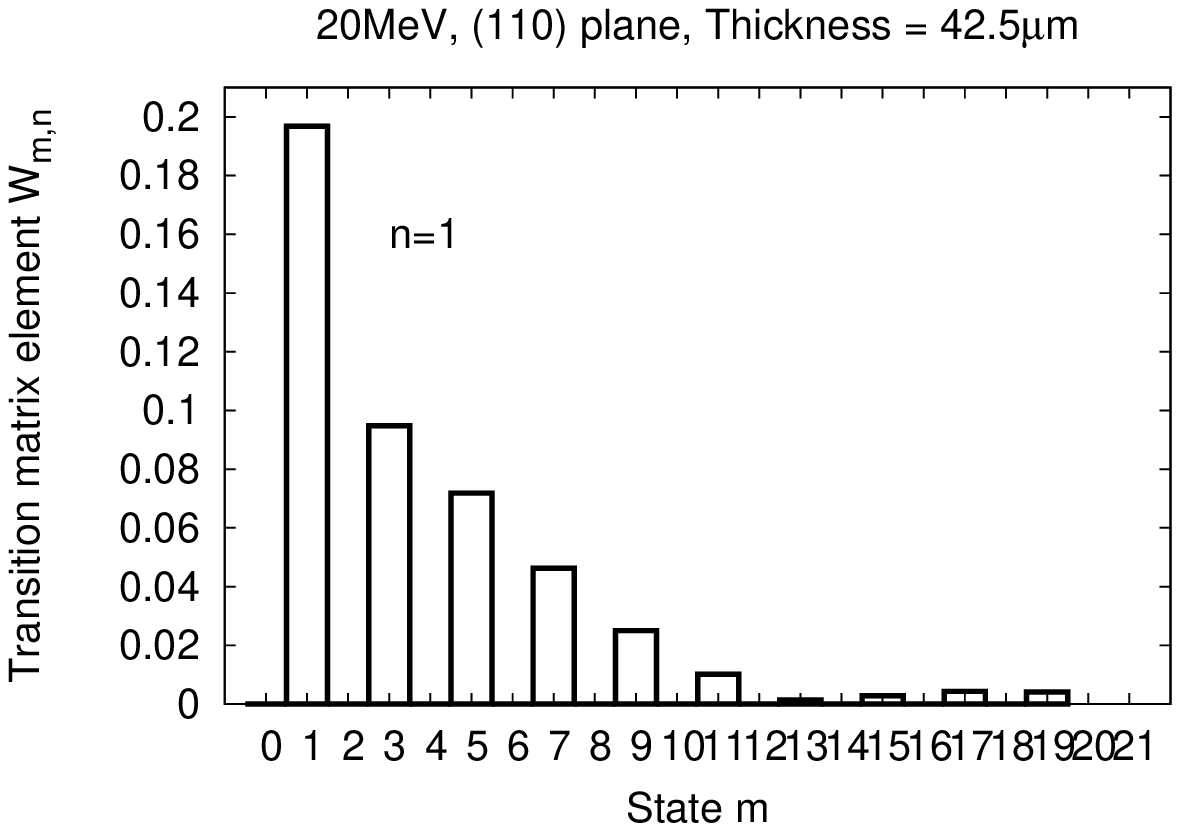}
\includegraphics[scale=0.5]{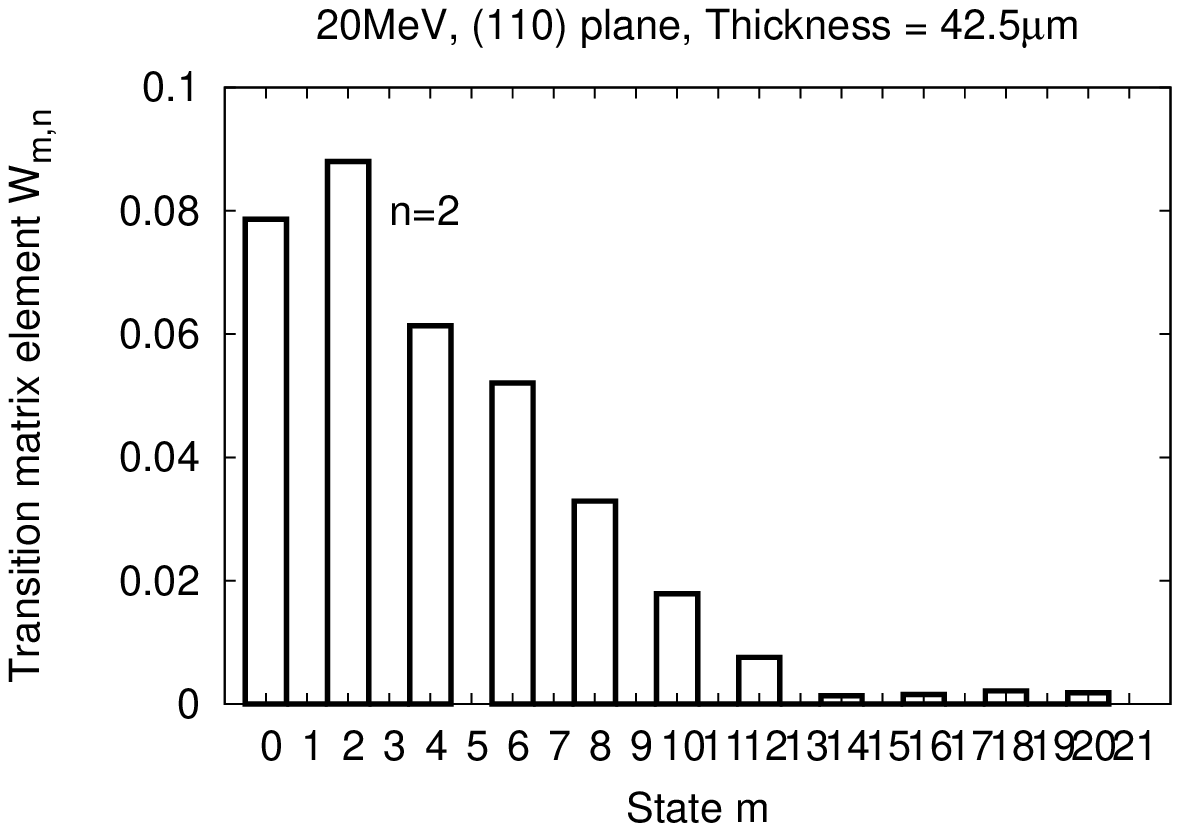}
\includegraphics[scale=0.5]{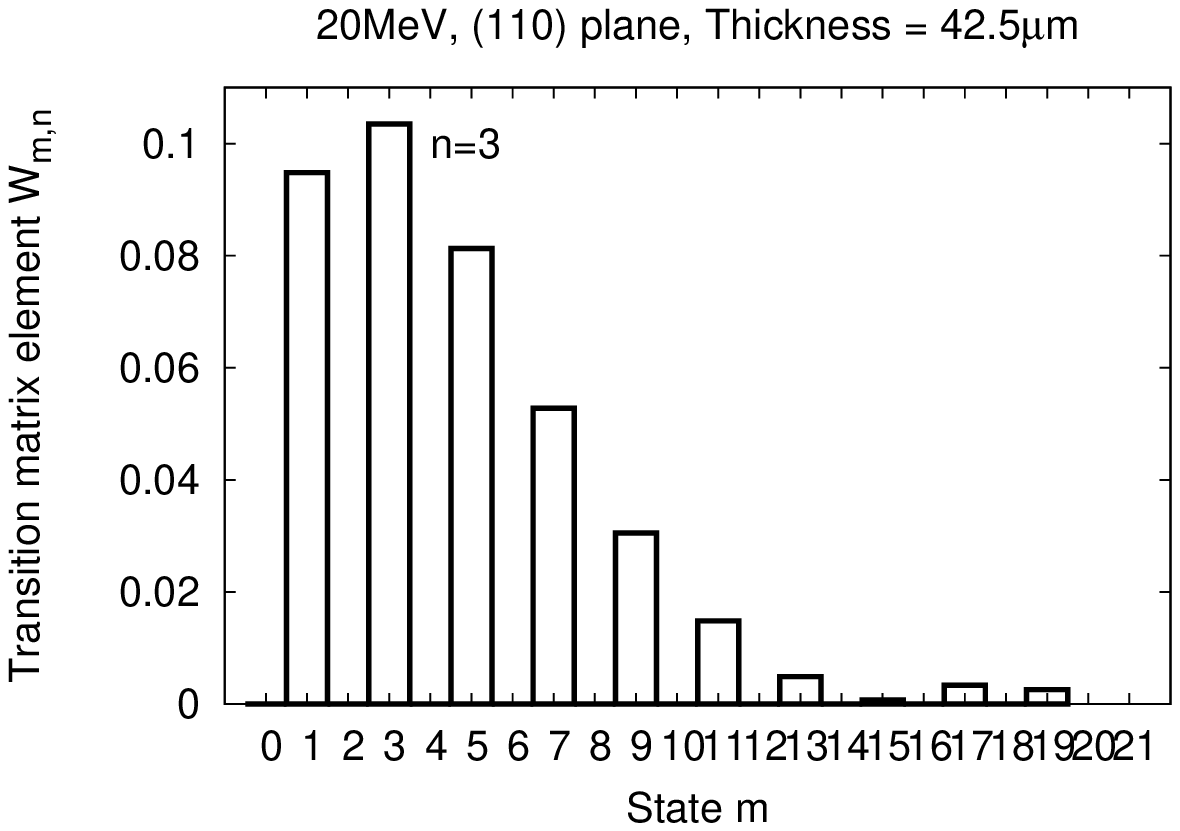}
\caption{Transition probabilities for the non-radiative transitions
due to thermal scattering from  the four bound states at the beam energy of 
20 MeV. }
\label{fig: tprob_1}
\end{figure}

Figure \ref{fig: tprob_1} shows the transition probabilities $W_{mn}$
calculated using Eqs. (\ref{eq: Wmn}) and (\ref{eq: Vimag}) for the 
four bound states at 20 MeV. As mentioned earlier, they obey the approximate
selection rule $W_{mn}=0$ if $|m-n|= odd$. The diagonal matrix element
$W_{nn}$ is the largest for each $n$ and decreases with increasing 
energy transfer as $|m-n|$ increases. Several conclusions can be drawn from
these transition rates. For example, most of the transitions from the
lower bound states are to other bound states. 
At $n=0$, only 16\% of the transitions take an electron to a free state 
$n \ge 4$, this increases to 36\% from the next bound state $n=1$ and 
to 52\% from $n=2$. Since the transition rates $W_{nm}$ are larger at lower $n$,
the bound states will depopulate faster than the free states will be populated.

\begin{figure}[h]
\centering
\includegraphics[scale=0.55]{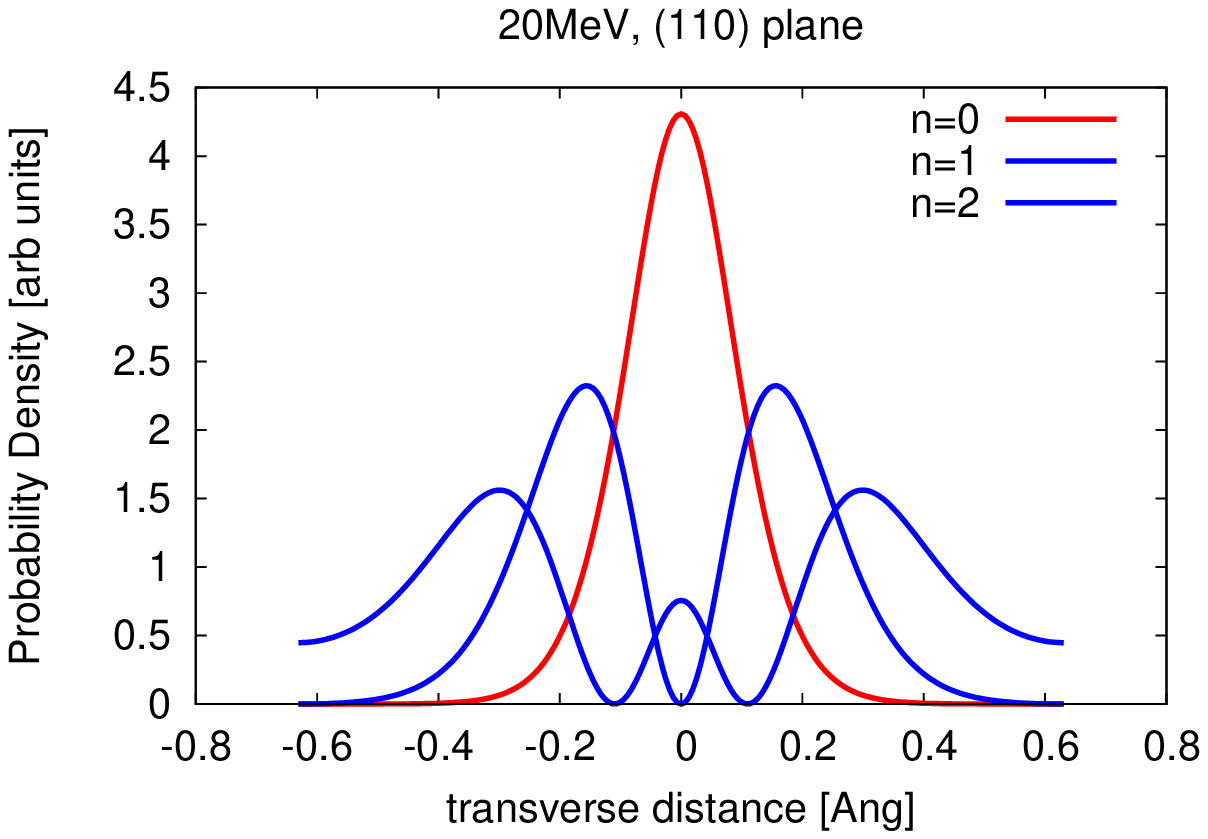}
\includegraphics[scale=0.55]{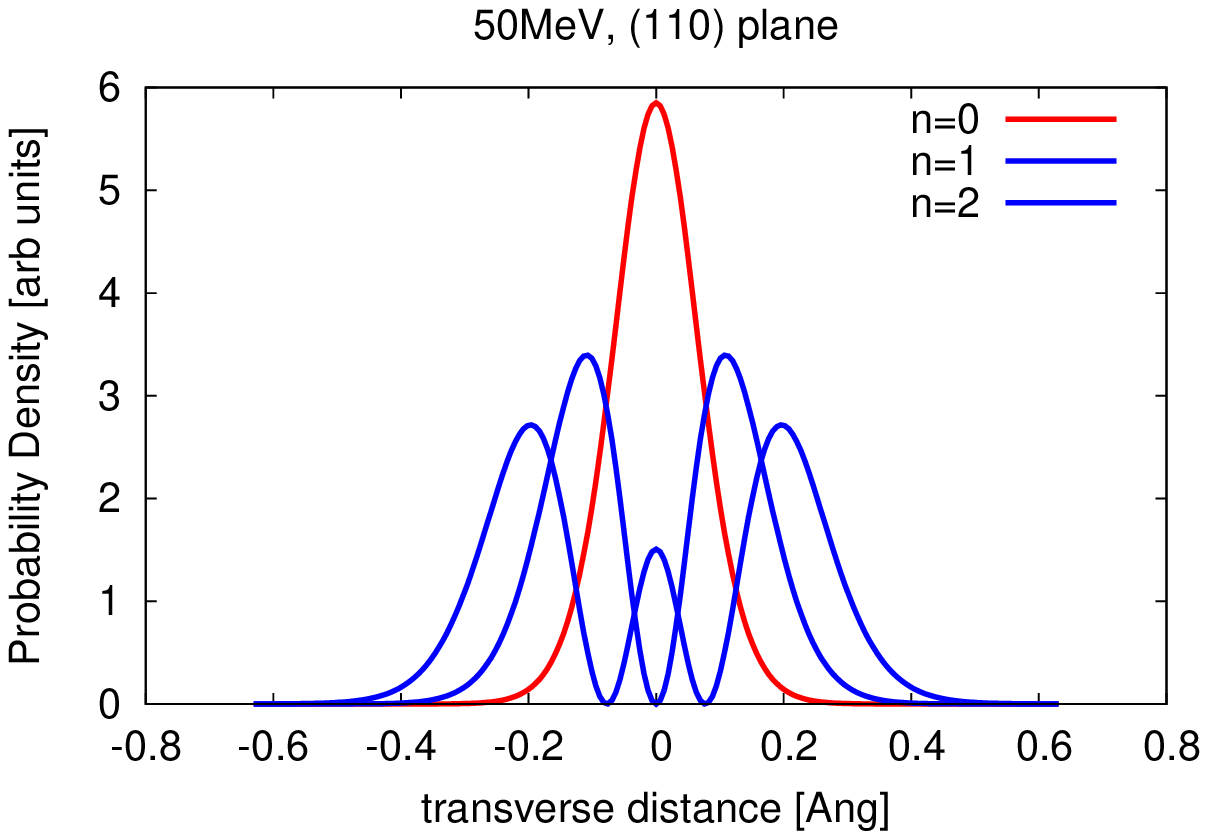}
\caption{Probability density as a function of the transverse distance
from the center of an atomic plane for the first three bound states.}
\label{fig: probden_20_50MeV}
\end{figure}

Figure \ref{fig: probden_20_50MeV} shows the probability density of the
first three bound states at beam energies of 20 MeV and 50 MeV. 
The eigenstates have definite parity, consequently the even states have a local
maximum at the nucleus while the odd states have a node at the
nucleus. We also observe that the probability densities of these states increase 
slightly with energy and they are more localized around the nucleus at 50 MeV.

\begin{figure}[h]
\centering
\includegraphics[scale=0.55]{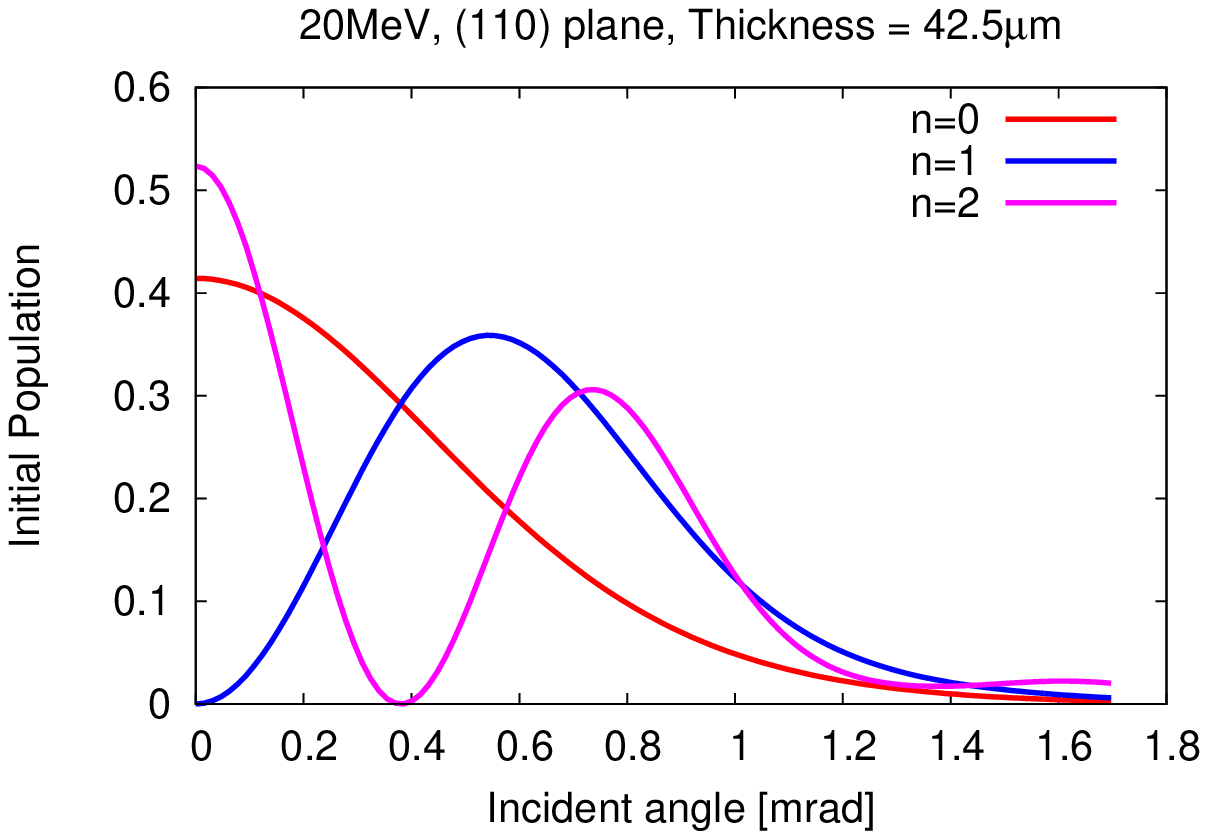}
\includegraphics[scale=0.55]{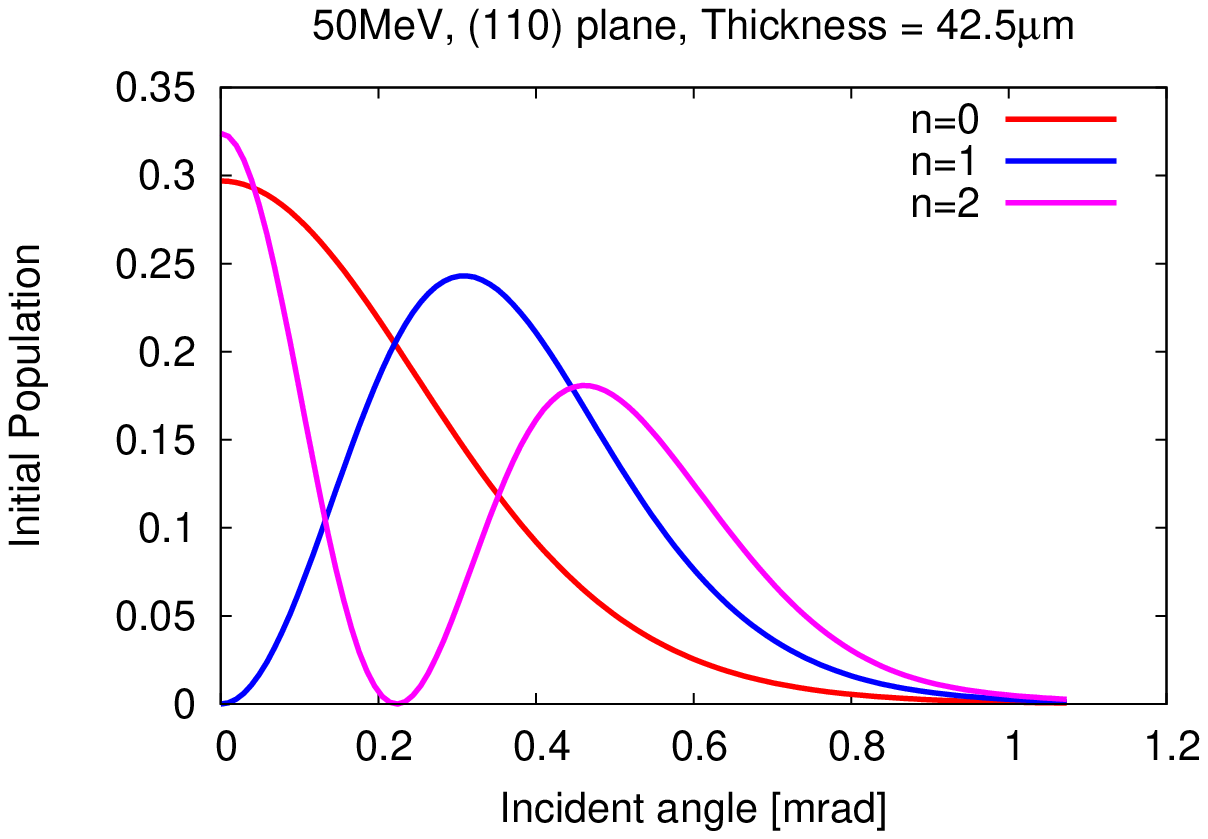}
\caption{Initial population as a function of the incident angle for the
3 lowest bound states. Left: At electron energy of 20 MeV. The maximum in the 
$n=1$ state occurs at an angle of 0.54 mrad. Right: At 50 MeV. The maximum in the 
$n=1$ state occurs at an angle of 0.3 mrad.}
\label{fig: initpop_ASTA_20_50}
% \end{figure}
% \begin{figure}
\centering
\includegraphics[scale=0.55]{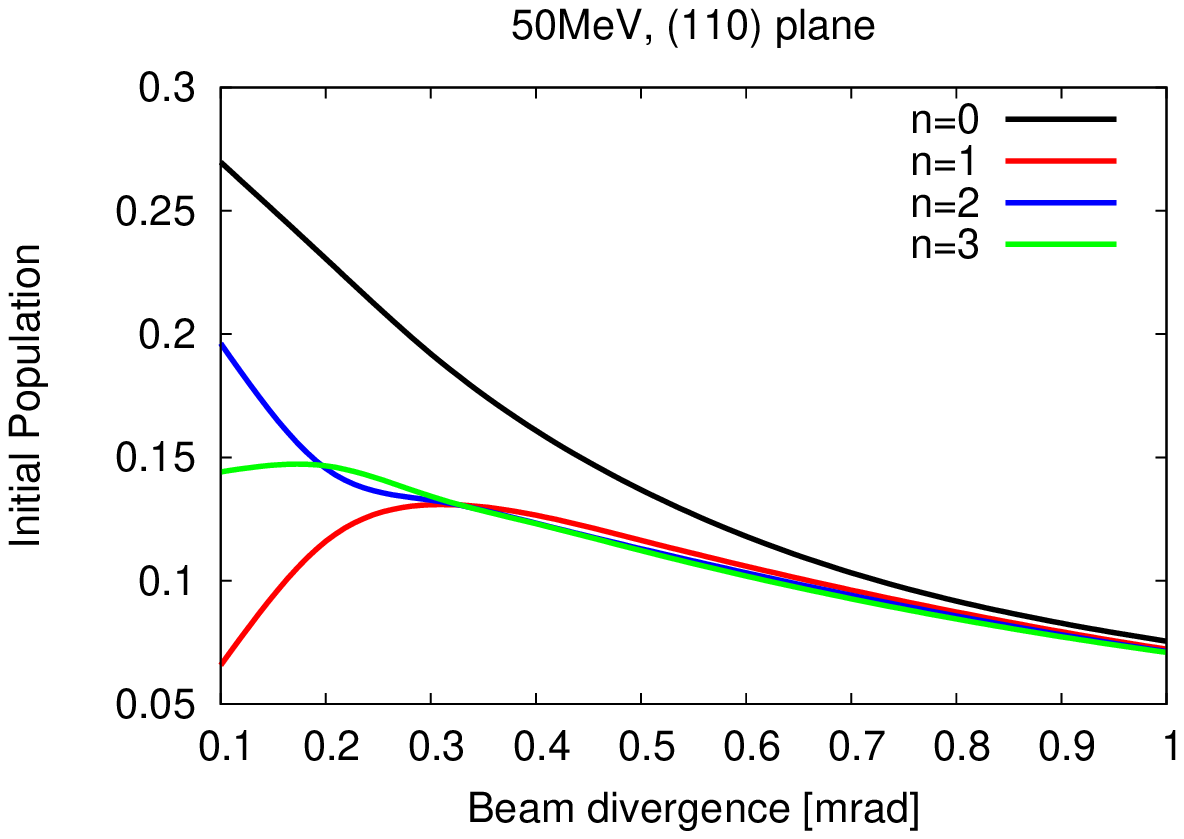}
\includegraphics[scale=0.55]{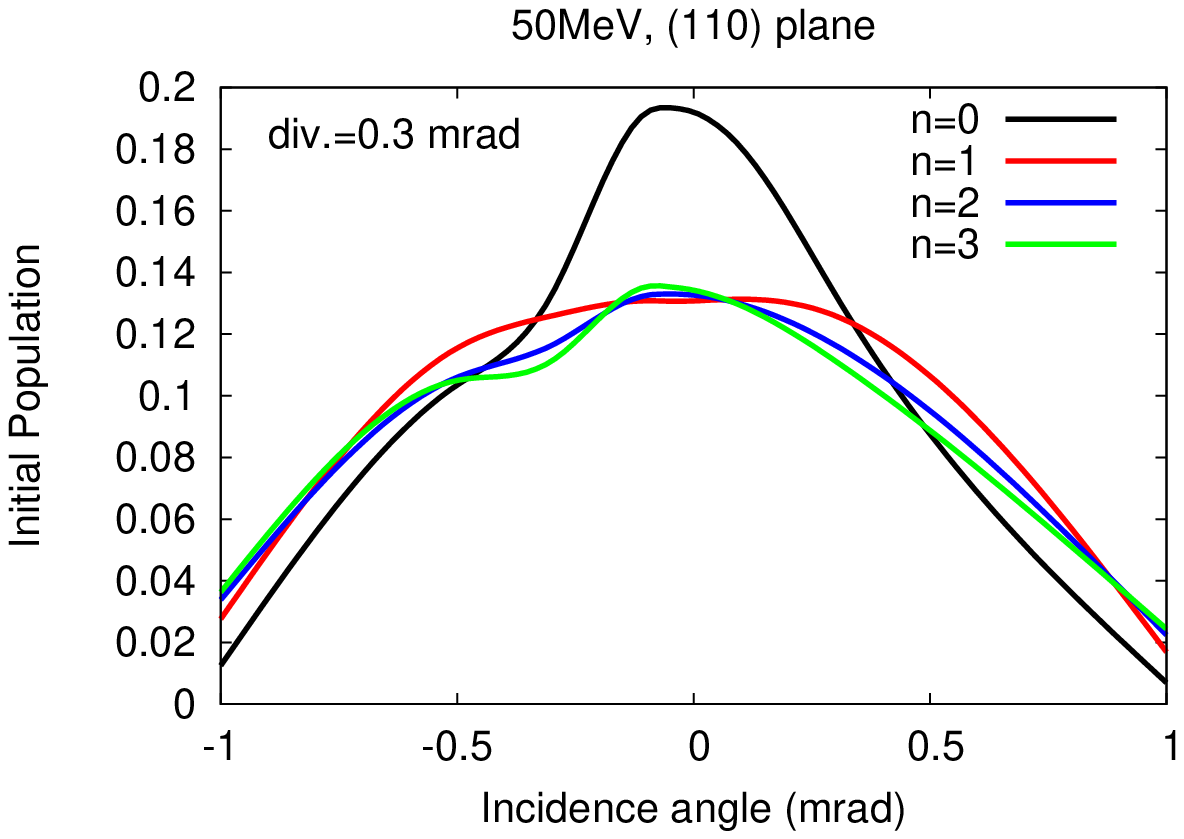}
\caption{Left: Initial populations in the lowest four bound states as a function
of the beam divergence at a beam energy of 50 MeV. The population in the
$n=1$ state has a maximum at a divergence of 0.3 mrad. Right: Populations in the
same states as a function of the incident angle at a beam divergence of 0.3 mrad
and energy 50 MeV.}
\label{fig: initpops_div_50}
\end{figure}

Figure \ref{fig: initpop_ASTA_20_50} shows the initial population at the 
entrance of the crystal for different incidence angles. The transverse energy
increases with incidence angle and the initial populations in these states
also change, in particular being non-zero for the odd states as well. Increased
initial population $P_1(0)$ in the $n=1$ state would increase the photon flux 
in the 1$\rarw$0 transitions. At 20 MeV, $P_1(0)$ is maximum at an incidence
angle of 0.54 mrad while at 50 MeV, the maximum is at 0.3 mrad.
The left plot in Fig.\ref{fig: initpops_div_50} shows the initial population 
as a function of the beam divergence at beam energy of 50 MeV. These 
populations will be dominated by the electrons incident at close to zero angle.
Thus in the even states we observe a slow decrease with divergence and not
the oscillations seen in the higher even states in 
Fig. \ref{fig: initpop_ASTA_20_50}. In the odd states however,
the non-zero contributions are due to electrons with non-zero incident angle
and thus in the $n=1$ state we observe a slow rise and a broad maximum at a
beam divergence of 0.3 mrad, matching the maximum location seen in Fig. 
\ref{fig: initpop_ASTA_20_50}. This optimum divergence is well below the
critical angle 0.98 mrad for channeling at 50 MeV. The right plot in Fig.
\ref{fig: initpops_div_50} shows the initial populations as a function of 
the incidence angle when the beam divergence is set to 0.3 mrad to maximize
the population in the $n=1$ state. Now we observe that the maximum in all
states is obtained at zero incidence angle, so there is no advantage in
tilting the crystal with respect to the beam direction when the beam
divergence is optimum. The same observations hold at 20 MeV where the optimum
beam divergence is about 0.5 mrad.

From the decay of the populations with distance into the crystal, we find
that the occupation lengths are about 9 $\mu$m with a 42.5 $\mu$m thick crystal
and about 20 $\mu$m with a 168 $\mu$m thick crystal. These values are about the 
same at 20 and 50 MeV and for the different bound states. 

%\clearpage

\subsection{X-ray energies, linewidths, photon yields} \label{subsec: EWY_ASTA}

\begin{figure}
\centering
\includegraphics[scale=0.55]{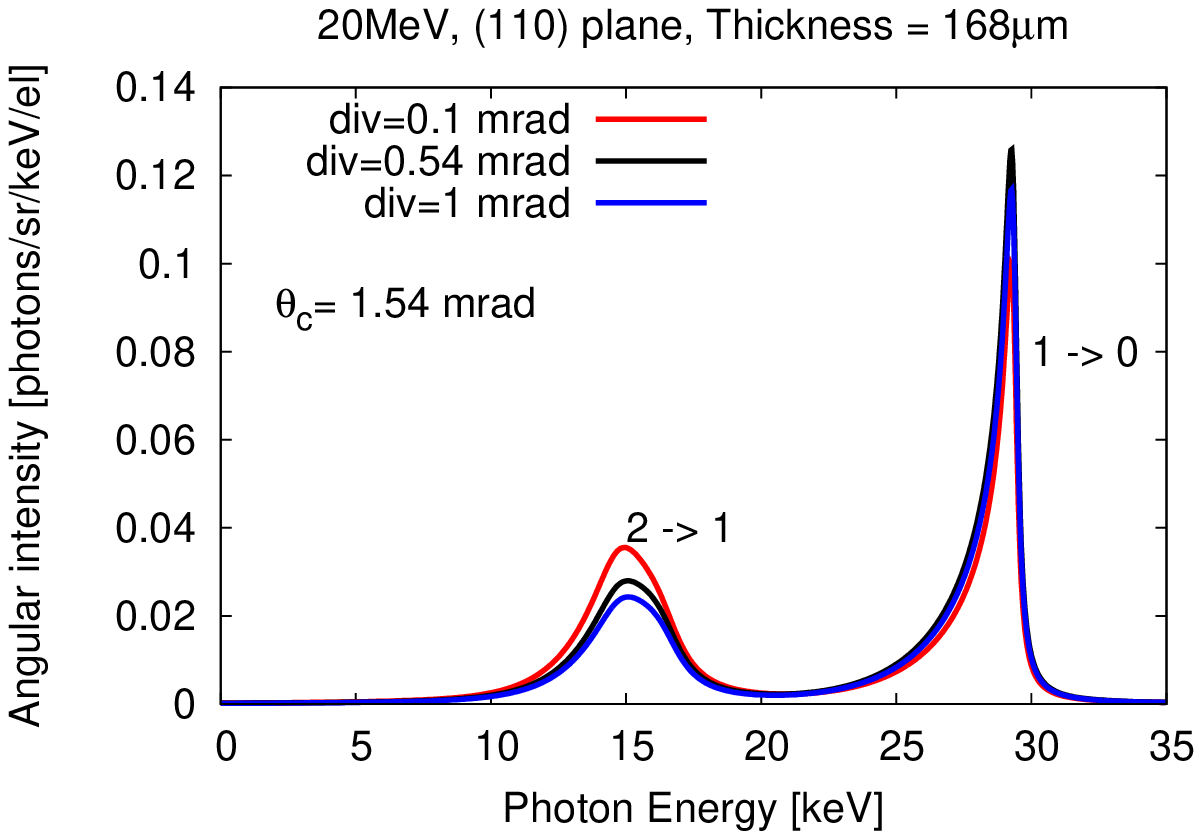}
\includegraphics[scale=0.55]{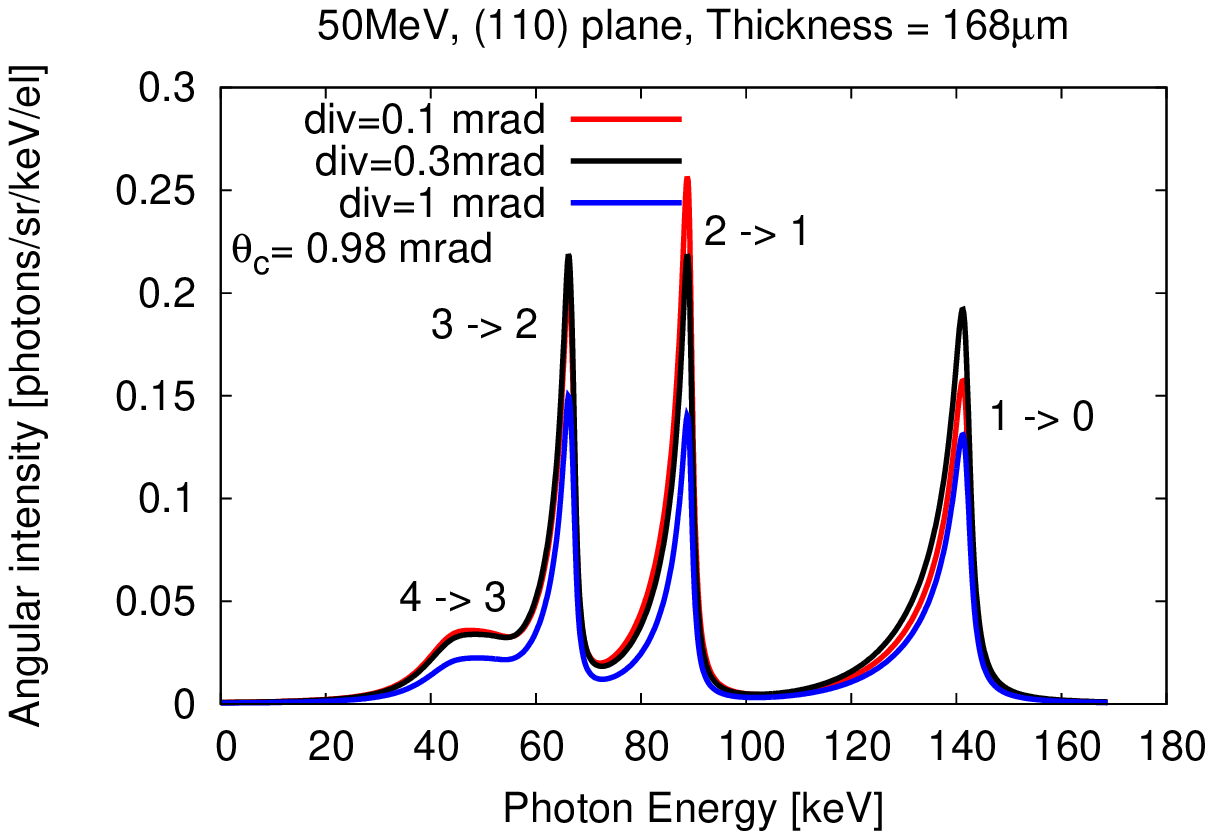}
\caption{Angular intensity spectrum at three values of the beam divergence 
with crystal thickness of 168 $\mu$m Left: 20 MeV, Right: 50 MeV}
\label{fig: intens_ASTA_div_2}
\end{figure}

We discuss the X-ray intensity spectrum expected at ASTA and consider the
effects of beam divergence on the spectrum. 
Fig. \ref{fig: intens_ASTA_div_2} 
shows the angular intensity spectrum (photons/sr-electron) with different
beam divergences for a crystal thickness of 168 $\mu$m at two energies. 
At the beam energy of 20 MeV, the 1$\rarw$0 transition leads to the highest
peak at 29.3 keV with a width of 1.8 keV while the 2$\rarw$1 transition leads
to a lower peak at 16.5 keV with a broader width of about 2.1 keV. From our
discussion of the ELBE simulations, we expect these linewidths to 
under-estimate the experimental width by roughly a factor of two. 

At a beam energy of 50 MeV there are more bound states and we observe more 
lines in the spectrum. The highest energy peak is still from the 1$\rarw 0$
transition
at 141.9 keV with a width of 9 keV while the most intense peak is from the
2$\rarw$1 transition at 89.3 keV with a width of 5.7 keV. There are also lower
energy and less intense lines
from the 3$\rarw$2 transition at 66.3 keV  and from the 4$\rarw$3 transition at
53.6 keV. 
In these calculations, the effect of the beam divergence on
the initial populations in the different states is included but not the change
of channeling fraction with the divergence. 
The spectrum with a beam divergence of 0.1 mrad is very close
to that of the single electron spectrum for both energies. 

At 20 MeV, the yield in the 1$\rarw$0 transition at 0.54 mrad divergence is 
higher compared to the yield at 0.1 mrad, but decreases on further increasing
the divergence to 1 mrad.
At 50 MeV, similar behavior is observed with the maximum in the 1$\rarw$0
and the 3$\rarw$2 transitions at a divergence of 0.3 mrad. Since the divergence
affects the dechanneling fraction, the observed spectrum may have a somewhat
different dependence on the beam divergence. 

\begin{table}
\caption{Expected X-ray energies, linewidths and photon yields with and without
dechanneling. The yields were calculated with the beam divergence set to 0.1 mrad and the incidence angle to zero. .At 50 MeV, the values for both the 
1$\rarw$0 and 2$\rarw 1$ transitions are shown.}
{\btable{|c|c|c|c|c|c|} \hline
$e^-$ Energy  & Thickness & Energy  & 
Linewidth  & \multicolumn{2}{|c|}{Yield [phot/e-/sr]} \\
\mbox{} [MeV]  & $\mu$m & $E_{sim}$[keV] 
& $\Gm_{sim}$[keV] & No dechan. yield & $n_f$ $|$  yield \\
\hline
20.0 & 42.5 & 29.3 &  1.21 &  0.27  & $n_B+1$ $|$ 0.075 \\
     & 168  & 29.3 &  1.85  & 0.77 & $n_B+7$ $|$ 0.17 \\
50.0 & 42.5 & 89.3 & 3.83  & 2.7 & $n_B+1$ $|$ 1.00 \\
     & 42.5 & 141.9 & 6.1  & 2.1 & $n_B+1$ $|$ 0.65 \\
     & 168 &  89.3 & 5.65  & 6.6  & $n_B+7$ $|$ 1.7 \\
     & 168 &  141.9 & 8.96  & 6.1  & $n_B+7$ $|$ 1.6 \\
\hline
\etable \label{table: ASTA_EWY} }
\end{table}
Table \ref{table: ASTA_EWY} shows the X-ray energies, linewidths and
photon yields expected at ASTA. Based on the ELBE simulations, the energies are 
expected to be accurate to better than 10\%. However the linewidths will 
be about a factor of two larger than the values in this table, as follows
from the discussion in Section \ref{sect: ELBE_EWY}. This table also shows
the photon yields for two cases: without enhanced dechanneling and with this dechanneling
with the parameter $n_f$ set to the value which under-estimates the
yield; see the discussion following Table \ref{table: ELBE_yields} and Fig.
\ref{fig: elbe_compare}.

%\clearpage

\subsection{Spectral brilliance}

A radiation source is usually characterized by 
the number of photons emitted per second per bandwidth per unit solid angle and
 unit area of the source, also called the spectral brilliance. 
The photon yields found above can be used to estimate the expected X-ray 
spectral brilliance at ASTA. 
The yield as calculated in Section \ref{subsec: EWY_ASTA} depends on the 
beam divergence through the dependence of the initial population on the 
divergence, as shown in Fig. \ref{fig: initpops_div_50}. It does not include
the likelihood that particles in the distribution with incidence angles greater
than the Lindhard critical angle $\theta_C$ will not be channeled. 
With the assumption of no rechanneling, the yield could be multiplied by the
fraction of particles with incident angles less than $\theta_C$,
\beq
f(|\theta| \le \theta_C) = \frac{1}{\sqrt{2\pi}\sg_e^{'}}
\int_{-\theta_C}^{\theta_C}
\exp[- \frac{\theta^2}{2(\sg_e^{'})^2}] d\theta = 
{\rm Erf}[\frac{\theta_C}{\sqrt{2}\sg_e^{'}}]
\eeq
where Erf is the error function, $\sg_e^{'}$ is the electron beam 
divergence and $\theta_C=\sqrt{2U_0/E_e}/\beta$ for
planar channeling with $U_0$ the depth of the potential. 

The average brilliance of the radiation emitted by a beam of electrons can be
written in terms of the differential intensity spectrum per electron as
\beq
B_{av} = \frac{d^2 N}{d\om d\Om}\frac{I_{av}}{e}\frac{E_{\gm}}{(\sg_{\gm})^2}
{\rm Erf}[\frac{\theta_C}{\sqrt{2}\sg_e^{'}}]
\eeq
where $I_{av}$ is the average electron beam current, $E_{\gm}$ is the energy
of the X-ray line and $\sg_{\gm}$
is the X-ray beam spot size. Expressed in
terms of the yield per electron and in a 0.1\% band-width
 we have the average brilliance expressed in typical light source units
\beqr
B_{av} & = & \frac{I_{av}}{e} \frac{Y *10^{-3}}{(\sg_{\gm} \sg_{\gm}^{'})^2 
\Dl E_{\gm}/E_{\gm}} {\rm Erf}[\frac{\theta_C}{\sqrt{2}\sg_e^{'}}]
= \frac{I_{av}}{e} \frac{\gm^2 Y (\sg_e^{'})^2 10^{-3}}
{\eps_N^2 \Dl E_{\gm}/E_{\gm}} {\rm Erf}[\frac{\theta_C}{\sqrt{2}\sg_e^{'}}]
 \nonumber \\
&  &  \mbox{} \hspace{3cm} \;\;\; {\rm photons/s-(mm-mrad)^2- 0.1\% BW}
\label{eq: Brill_av}
\eeqr
$Y$ is the total  photon yield per electron, $\Dl E_{\gm}/E_{\gm}$ is the 
relative width of the X-ray line, and $\eps_N$ is the normalized  
emittance in mm-mrad.  We set the X-ray beam spot size to the
lower limit value of the electron beam spot size
 $\sg_{\gm} = \sg_e = \eps_N/(\gm \sg_e^{'})$,  while the X-ray divergence is 
$\sg_{\gm}^{'} = 1/\gm$.

\begin{figure}
\centering
\includegraphics[scale=0.55]{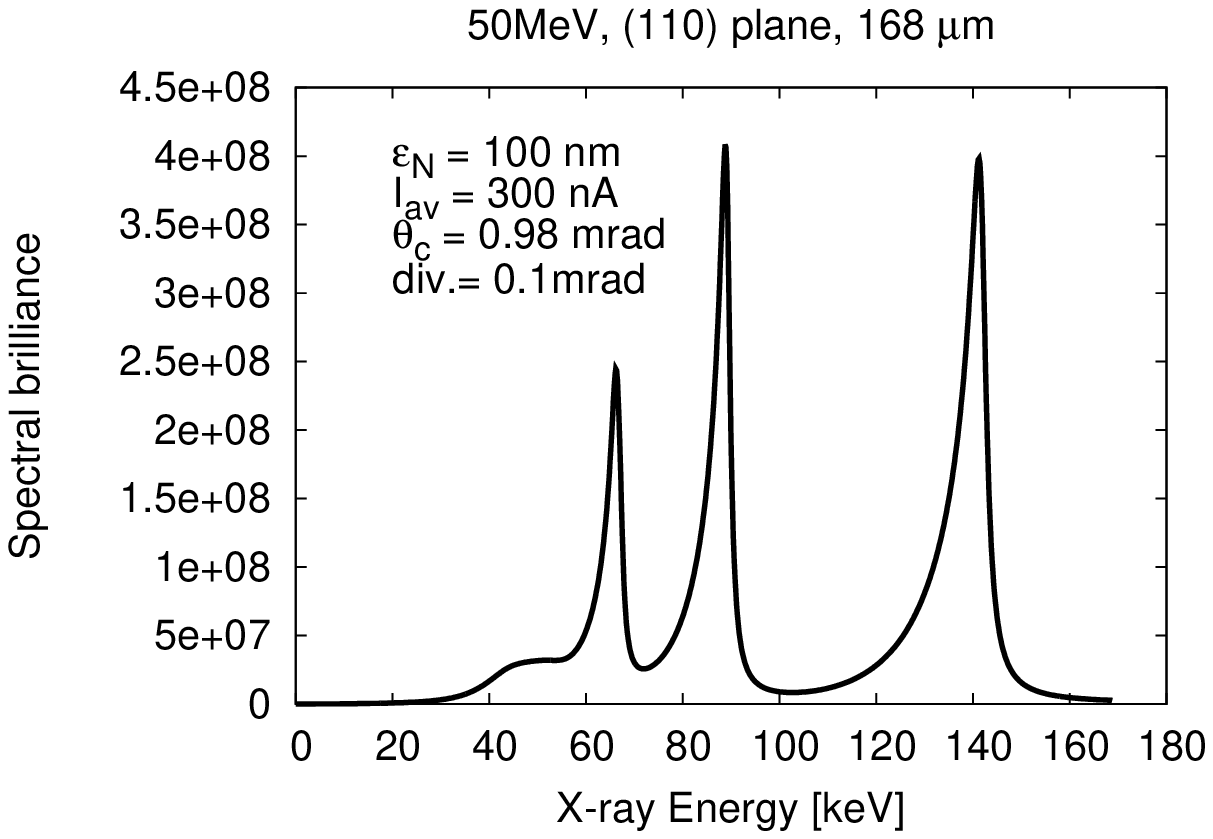}
\includegraphics[scale=0.55]{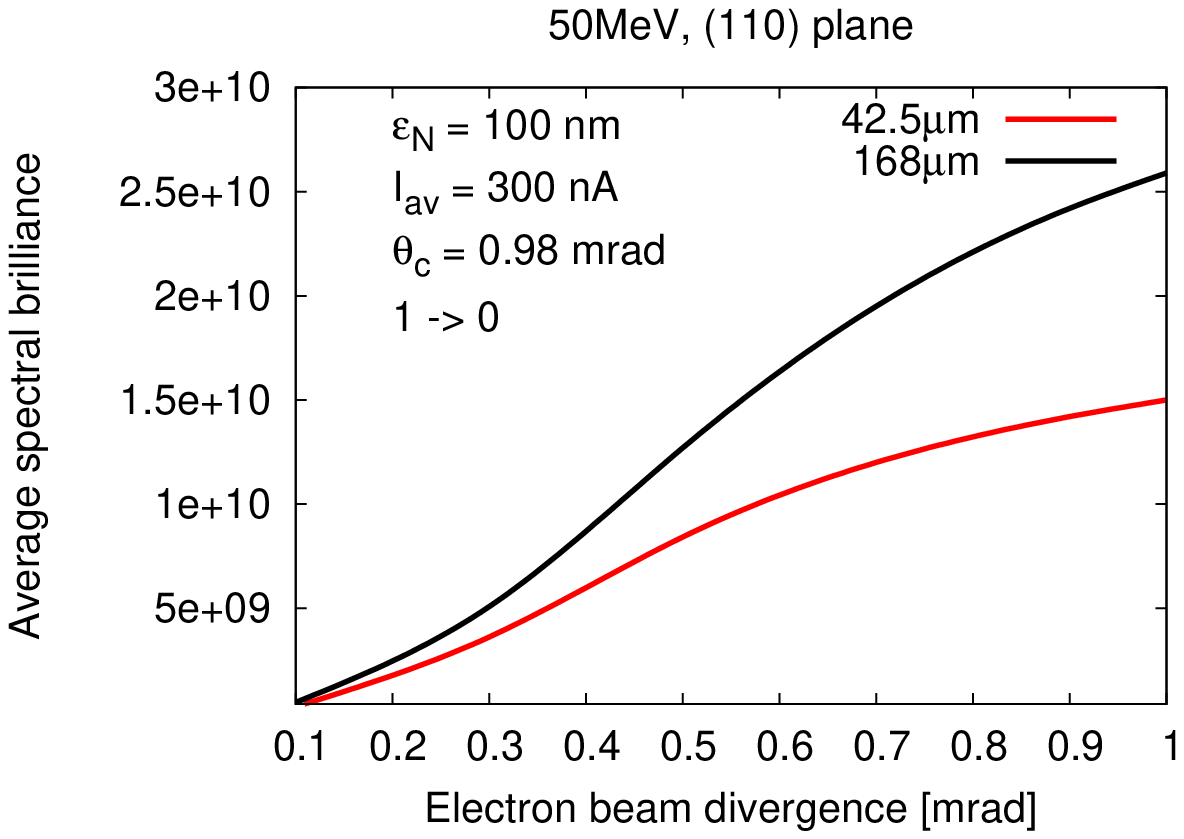}
\caption{Left: Average spectral brilliance of X-rays with 50 MeV electron 
beams as a function of the X-ray energy with beam divergence =0.1 mrad.
Right: Spectral brilliance as a function of the beam divergence for the
 1$\rarw 0$ transition at two values of the crystal thickness.}
\label{fig: brilliance_50MeV}
\end{figure}
The left plot in Fig.~\ref{fig: brilliance_50MeV} shows the brilliance as
a function of X-ray energy at a beam energy of 50 MeV with a beam divergence of
0.1 mrad while the
right plot in this figure shows the expected brilliance in the 1$\rarw$0 
line as a function of the beam divergence for two crystal thicknesses.
With the assumptions made above, we observe that
that the brilliance is larger with the thicker crystal, the difference
increases with divergence and reaches about 70\%
when the beam divergence equals the Lindhard critical angle.

Table \ref{table: ASTA_brilliance} shows the brilliance and photon flux at two
 energies for a crystal thickness of 168 $\mu$m, again with the same assumptions
as in Figure \ref{fig: brilliance_50MeV}. Since the values quoted are for the
beam divergence of 0.1 mrad, the value quoted for the ELBE experiment and 
used in setting the value of $n_f$ in Eqs. (\ref{eq: Pn_mod_1}) and
(\ref{eq: Pn_mod_2}), the deviations
from the values to be observed at ASTA may be small. 
\begin{table}
\caption{X-ray brilliance and photon flux from the 1$\rarw 0$ transition with ASTA
parameters for two energies and crystal thickness of 168 $\mu$m. 
The estimated energy spreads shown are a factor of two larger than the 
calculated values. $\ddag$ Units of
the brilliance are : photons/(s-(mm-mrad)$^2$-0.1\% BW)} 
{ \btable{|c|c|c|} \hline
Beam energy [MeV] & 20 &  50 \\
Av. beam current [nA] & 300 & 300 \\
Beam emittance [nm] & 100 & 100 \\
Beam divergence [mrad] & 0.1 & 0.1 \\
X-ray energy from $1\rarw 0$ $E_{\gm}$ [keV] & 29.2 & 141.9  \\
Est. energy spread $\Dl E_{\gm}$ [keV] & $\sim 4$  & $\sim 18$ \\
Angular yield [photons/(e-sr)] & 0.17  &  1.69    \\
Absolute yield/electron [$\times 10^{-3}$] & 0.11  &  0.17   \\
Av. X-ray brilliance  [$\times 10^{7}$] $\ddag$ &  0.79 &  48.0    \\
Av. Photon flux at $E_{\gm}$ [photons/s] $\times 10^8$ & 2.1  & 3.3  \\
\hline
\etable \label{table: ASTA_brilliance} }
\end{table}

As steps towards increasing the brilliance, one could consider increasing the
beam current either with a higher bunch charge or a higher micropulse repetition
rate if the crystal does not suffer damage from heating at the higher currents.
A more promising path would be to lower the emittance since the brilliance
depends inversely on the square of the emittance. 
The results above have assumed an electron emittance of
100 nm using a laser photocathode.  First tests of
operation with a field emission cathode mentioned above have recently been 
reported \cite{Piot_14}.
Assuming that success is achieved with these cathodes and that the low emittance
generated can be preserved until the crystal, the brilliance could then be
increased by about two orders of magnitude above the values reported here.

%\clearpage

\section{Conclusions}

In this report we have studied the expected spectral brilliance of X-rays from
channeling experiments to be performed at the ASTA photoinjector. We revisited
the theoretical model, corrected the potential describing thermal scattering
and developed a heuristic model to include dechanneling in the population
dynamics. We used the updated 
model to first compare with the experimental values reported from the ELBE
facility and second to predict values for ASTA.

We compared the energies, linewidths and photon yields from the model with
the results at the ELBE facility. With appropriate choices of
dechanneling states in the model, the simulated yield agrees well with observed
photon yields, see Fig.\ref{fig: elbe_compare}.
The theoretical linewidth is about a factor of two smaller than the
observed values. This is due to the neglect of electron scattering with
the atomic electrons and the plasmonic modes. This scattering affects only the
linewidth but does not 
affect the photon yields. From the population dynamics we were able to
estimate, for different quantum states, the occupation length whose classical 
analog is the dechanneling length. The occupation length was found to increase
with crystal thickness but was nearly independent of beam energy in the 
energy range studied. This pointed to the importance of rechanneling in the
quantum regime where particles in the free states can be scattered
back into the channeled bound states. Rechanneling increases with crystal
thickness and explains why the measured occupation lengths are longer than
simple classical estimates. We found that 
the optimum crystal thickness to maximize the intensity of the 1$\rarw$0
transition is about 7 times the occupation length.

When applied to ASTA, the model finds that with an electron beam energy of 50
MeV, X-ray peaks are expected at about 142 keV from the 1$\rarw$0 transition and
at 89 keV from the 2$\rarw$1 transition with linewidths around 14\%. The ability
of channeling radiation to produce hard X-rays with moderate beam energies is 
one of the main premises for these experiments. We find that with a crystal 
thickness of 168 $\mu$m and 
electron transverse emittances of 100 nm and beam current of 300 nA, the 
expected brilliance is of the order of 10$^{10}$ photons/(s-(mm-mrad)$^2$-0.1\%
BW). It is possible that thicker crystals may increase the brilliance 
above these values. Significant increase in the brilliance by about two
orders of magnitude could be achieved with ultra-low emittance beams using 
field emitter cathodes and
beam studies with these novel cathodes are in progress.

\vspace{2em}

\noi {\bf \large Acknowledgments} \newline
We thank the Lee Teng undergraduate internship program at Fermilab which awarded 
C. Lynn a summer internship in 2013 when this project began. We thank B.~Blomberg,
D.~Mihalcea and P.~Piot for useful discussions. 
Fermilab is operated by Fermi Research Alliance, LLC under Contract 
No. DE-AC02-07CH11359 with the United States Department of Energy.

%\clearpage

\end{document}